\documentclass[pra,a4paper,twocolumn,showpacs,floatfix,superscriptaddress,amsmath,preprintnumbers,citeautoscript,aps,longbibliography]{revtex4-1}
\pdfoutput=1
\usepackage[T1]{fontenc}\usepackage[latin1]{inputenc}
\usepackage{dcolumn,graphicx,color,booktabs,microtype,afterpage} \graphicspath{{Figures/}}
\usepackage[charter,greekuppercase=italicized]{mathdesign}
\renewcommand{\figurename}{Fig.}
\renewcommand{\tablename}{Table}
\makeatletter\renewcommand{\fnum@figure}[1]{\figurename~\thefigure.}\makeatother
\makeatletter\renewcommand{\fnum@table}[1]{\tablename~\thetable.}\makeatother
\newcount\hh \newcount\mm
\hh=\time \divide\hh by 60
\mm=\hh \multiply\mm by 60 \mm=-\mm
\advance\mm by \time
\def\now{\number\hh:\ifnum\mm<10{}0\fi\number\mm}
\usepackage[colorlinks,plainpages=false,linkcolor=blue,urlcolor=blue,citecolor=blue,pdfpagemode=UseNone,pdfstartview=FitBH]{hyperref}

\newcommand{\TN}{$T_{\mathrm{N}}$}

\begin{document}
\makeatletter\renewcommand{\ps@plain}{%
\def\@evenhead{\hfill\itshape\rightmark}%
\def\@oddhead{\itshape\leftmark\hfill}%
\renewcommand{\@evenfoot}{\hfill\small{--~\thepage~--}\hfill}%
\renewcommand{\@oddfoot}{\hfill\small{--~\thepage~--}\hfill}%
}\makeatother\pagestyle{plain}

\title{Gradual pressure-induced enhancement of magnon excitations in CeCoSi}

\author{S.~E.~Nikitin}
\email{nikitin@cpfs.mpg.de}
\affiliation{Max Planck Institute for Chemical Physics of Solids, D-01187 Dresden, Germany}
\affiliation{Institut f{\"u}r Festk{\"o}rper- und Materialphysik, Technische Universit{\"a}t Dresden, D-01069 Dresden, Germany}
\author{D. G. Franco}
\affiliation{Max Planck Institute for Chemical Physics of Solids, D-01187 Dresden, Germany}
\affiliation{Centro At\'{o}mico Bariloche and Instituto Balseiro, Comisi\'{o}n Nacional de Energ\'{i}a At\'{o}mica (CNEA), Universidad Nacional de Cuyo, Consejo Nacional de Investigaciones Cient\'{i}ficas y T\'{e}cnicas (CONICET), Av. E. Bustillo 9500, R8402AGP San Carlos de Bariloche, R\'{i}o Negro, Argentina}
\author{J. Kwon}
\affiliation{Max Planck Institute for Chemical Physics of Solids, D-01187 Dresden, Germany}
\author{R. Bewley}
\affiliation{ISIS Facility, STFC Rutherford Appleton Laboratory, Harwell Campus, Didcot OX11 0QX, United Kingdom}
\author{A.~Podlesnyak}
\affiliation{Neutron Scattering Division, Oak Ridge National Laboratory, Oak Ridge, Tennessee 37831, USA}
\author{A. Hoser}
\affiliation{Helmholtz-Zentrum Berlin f\"ur Materialien und Energie, D-14109 Berlin, Germany}
\author{M. M. Koza}
\affiliation{Institut Laue-Langevin, F-38042 Grenoble Cedex 9, France}
\author{C. Geibel}
\affiliation{Max Planck Institute for Chemical Physics of Solids, D-01187 Dresden, Germany}
\author{O. Stockert}
\affiliation{Max Planck Institute for Chemical Physics of Solids, D-01187 Dresden, Germany}
\thanks{Corresponding author: nikitin@cpfs.mpg.de}

\begin{abstract}
CeCoSi is an intermetallic antiferromagnet with a very unusual temperature-pressure phase diagram: at ambient pressure it orders below \TN\ = 8.8~K, while application of hydrostatic pressure induces a new magnetically ordered phase with exceptionally high transition temperature of $\sim40$~K at 1.5~GPa. We studied the magnetic properties and the pressure-induced magnetic phase of CeCoSi by means of elastic and inelastic neutron scattering (INS) and heat capacity measurements. At ambient pressure CeCoSi orders into a simple commensurate AFM structure with a reduced ordered moment of only $m_{\mathrm{Ce}} = 0.37(6)$~$\mu_{\mathrm{B}}$. Specific heat and low-energy INS indicate a significant gap in the low-energy magnon excitation spectrum in the antiferromagnetic phase, with the CEF excitations located above $10$~meV. Hydrostatic pressure gradually shifts the energy of the magnon band towards higher energies, and the temperature dependence of the magnons measured at 1.5~GPa is consistent with the phase diagram. Moreover, the CEF excitations are also drastically modified under pressure.
\end{abstract}

\maketitle{}

\section{Introduction}

Ce-based intermetallic compounds represent a rich playground for exploration of quantum critical phenomena~\cite{stockert2011unconventional, gegenwart2008quantum, si2010heavy, stockert2011magnetically}. The ground state of these materials originates quite often from a competition between RKKY interaction and Kondo screening, which tend to create long-range magnetically ordered and nonmagnetic heavy-fermion states, respectively. The delicate balance between RKKY and Kondo effects can be quite easily tuned by an external tuning parameters, e.g. composition, uniaxial or hydrostatic pressure, magnetic field etc. Usually, application of hydrostatic pressure enhances the coupling between the conduction electrons and the localized Ce moments $J_{\mathrm{cf}}$, and therefore, drastically increases the strength of the Kondo effect ($T_{\mathrm{K}}\propto\mathrm{exp}(-\frac{1}{2J_{\mathrm{cf}}})$) leading to a reduced magnetic ordering temperature and shifting the ground state of the material closer towards a nonmagnetic heavy-fermion state~\cite{kitaoka1995nmr, lengyel2011pressure, coleman2015heavy, shen2020strange}. 

However, in several recent works it was shown that CeCoSi represents an intriguing counterexample to this paradigm~\cite{lengyel2013temperature, tanida2018substitution}. This material crystallizes in the tetragonal CeFeSi structure (space group $P4/nmm$) and the cerium moments order antiferromagnetically below the $T_{\mathrm{N}} = 8.8$~K~\cite{chevalier2004effect, chevalier2006antiferromagnetic}. Results of powder neutron diffraction measurements revealed a commensurate antiferromagnetic structure in isostructural CeCoGe  with a simple antiferromagnetic stacking of FM Ce planes along the $c$-axis~\cite{chevalier2004reinvestigation}, but the information about the magnetic structure of CeCoSi is absent to the best of our knowledge. Resistivity measurements under hydrostatic pressure~\cite{lengyel2013temperature} have shown that the application of rather moderate pressure of only $\sim$ 0.6~GPa induces a new magnetically ordered phase with exceptionally high transition temperature $T_{\mathrm{c}} \approx 40$~K (see the phase diagram in Fig.~\ref{PhaseDiagram}). The pressure-induced phase has a dome shape and the $T_{\mathrm{c}}$ changes only slightly up to $\sim 1.7$~GPa, whereas upon further pressure increase $T_{\mathrm{c}}$ gets rapidly suppressed and a quantum critical point, characterized by a divergence of resistivity parameters $A$ and $\rho_0$, was found at $\sim$~2.2~GPa~\cite{lengyel2013temperature}. A nonmagnetic Fermi-liquid state was observed at higher pressures.

In a recent study on single crystals, a very weak anomaly was observed in the specific heat and in the susceptibility at about 12~K and was proposed to be quadrupolar order~\cite{tanida2019successive}. Subsequent NMR and NQR results at high pressure indicate that the high-$T$ transition under pressure is a weak structural transition~\cite{Manago2019}. Its primary order parameter was also proposed to be an antiferroquadrupolar one. However Ce$^{3+}$ is a Kramers ion, and in solids its $J = 5/2$ multiplet is split into 3 Kramers doublets, which do not bear a quadrupolar degree of freedom. A quadrupolar order is then only possible by mixing excited CEF doublets, which requires the excited CEF states to be at low energy, of the order of the quadrupolar ordering temperature. However preliminary results indicated the CEF splitting to be much larger, larger than 100~K~\cite{lengyel2013temperature, tanida2019successive}, at least at ambient pressure. That would make a standard quadrupolar ordering not only at 12~K, but also at 35~K very unlikely. In order to clarify this question, reliable information on the CEF excitation energies is crucial.

\begin{figure}[b]	
\center{\includegraphics[width=.6\linewidth]{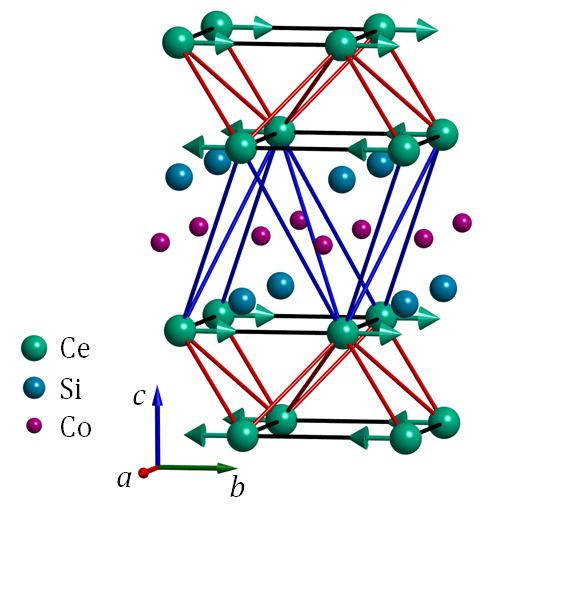}\vspace{-10pt}}
 		 \caption{~Sketch of crystal and magnetic structure of CeCoSi. Solid lines show the minimum set of three exchange interactions, which are needed to stabilize the magnetic ground state.}
  	\label{MagneticStructure}\vspace{-12pt}	
\end{figure}

It is worth noting that such a jump-like drastic increase of the transition temperature under the application of very moderate pressure is highly unusual for Ce-based metals and has no simple explanation in terms of RKKY/Kondo competition, and some authors~\cite{lengyel2013temperature} proposed a meta-orbital transition scenario to describe the appearance of a pressure-induced ordered phase. The concept of the meta-orbital transition was proposed by Kazumasa Hattori~\cite{hattori2010meta}. He investigated a two-orbital Anderson lattice model (orbital energy splitting is induced by the CEF effect) with Ising orbital intersite exchange interactions using a dynamical mean-field theory. 
It was shown, that if the hybridization between the ground-state $f$-electron orbital and conduction electrons is smaller than the one between the excited $f$-electron orbital and conduction electrons at low pressures, the occupancy of the two orbitals changes steeply upon application of pressure. In other words, the excited CEF excitations, which typically had been ignored, because in most cases the lowest excited CEF state is well separated to the ground state, may start to contribute to the ground state properties and induce the transition. Such a meta-orbital transition has been theoretically predicted to happen in CeCu$_2$Si$_2$~\cite{pourovskii2014theoretical}, but no experimental verification exists so far in any compound. Therefore, knowledge of the CEF splitting scheme, the magnon excitations and their pressure evolution can provide crucial information about the unusual physics of CeCoSi, which might be the first realization of a material exhibiting a meta-orbital transition.

To address these questions we synthesized polycrystalline samples of CeCoSi and its nonmagnetic counterpart LaCoSi. Then, we characterized the samples using neutron diffraction and specific heat measurements. The magnetic excitation spectra were investigated by means of elastic and inelastic neutron scattering under hydrostatic pressures up to 1.5~GPa.

\section{Experimental details}
The polycrystalline samples of CeCoSi and its nonmagnetic counterpart LaCoSi were synthesized from elemental Ce (La), Co and Si materials mixed in stoichiometric ratios using arc-melting technique, and then annealed for $\sim$2 weeks at a temperature close to 1200 $^{\circ}$C (the details are given in~\cite{lengyel2013temperature}). The resulting materials were examined using x-ray powder diffraction and energy dispersive x-ray spectroscopy analysis (EDX). The EDX measurements have shown that after the annealing, the majority of the sample consists of CeCoSi phase, with a small inclusion of an elemental Ce and CeCo$_2$Si$_2$ phase, but according to powder diffraction, the concentration of impurity phases is below 2~\%.

Neutron powder diffraction measurements were performed at the diffractometer E6 (HZB facility). The powder diffraction patterns were collected at $T = 1.7$ and 20~K with $\lambda = 2.41$~\AA. Inelastic neutron scattering (INS) measurements at ambient pressure were carried out at the time-of-flight (TOF) spectrometers IN4 and IN6 of the Institut Laue-Langevin in the temperature range 1.7--150~K. The incident neutron energies were fixed to $E_{\mathrm{i}} = 31.95$~meV and $E_{\mathrm{i}} = 3.86$~meV at IN4 and IN6 experiments, respectively. In these experiments we measured $\sim 10$~g of powder samples.

To study the effect of hydrostatic pressure on the spin excitations in CeCoSi we performed two INS experiments using the cold TOF spectrometers LET~\cite{bewley2011let} at ISIS neutron source and CNCS~\cite{CNCS1, CNCS2} at SNS, ORNL. In order to apply hydrostatic pressure in both experiments we used similar NiCrAl pressure cells with a relatively weak background in the inelastic channel and reasonable neutron transmission of $\sim$30~\% designed by Dr. Ravil Sadykov from the Institute for Nuclear Research, Moscow. The cells were filled with $\sim 1.5$~g of powder and fluorinert FC-770 was used as pressure transmitting medium.

At CNCS experiment we measured magnetic excitations with two neutron incident energies $E_{\mathrm{i}} = 6.15$~meV and $25.23$~meV to study magnon and CEF excitations, respectively. The measurements were performed at the base temperature of the orange cryostat, $T = 1.7$~K, and at three pressures of $P = 0.2, 0.6$ and 1~GPa. The pressure cell used in the CNCS experiment had an optical window, which allowed us to monitor the pressure by means of a ruby fluorescence method~\cite{forman1972pressure, podlesnyak2018clamp}. 

\begin{figure}[t]	
\center{\includegraphics[width=\linewidth]{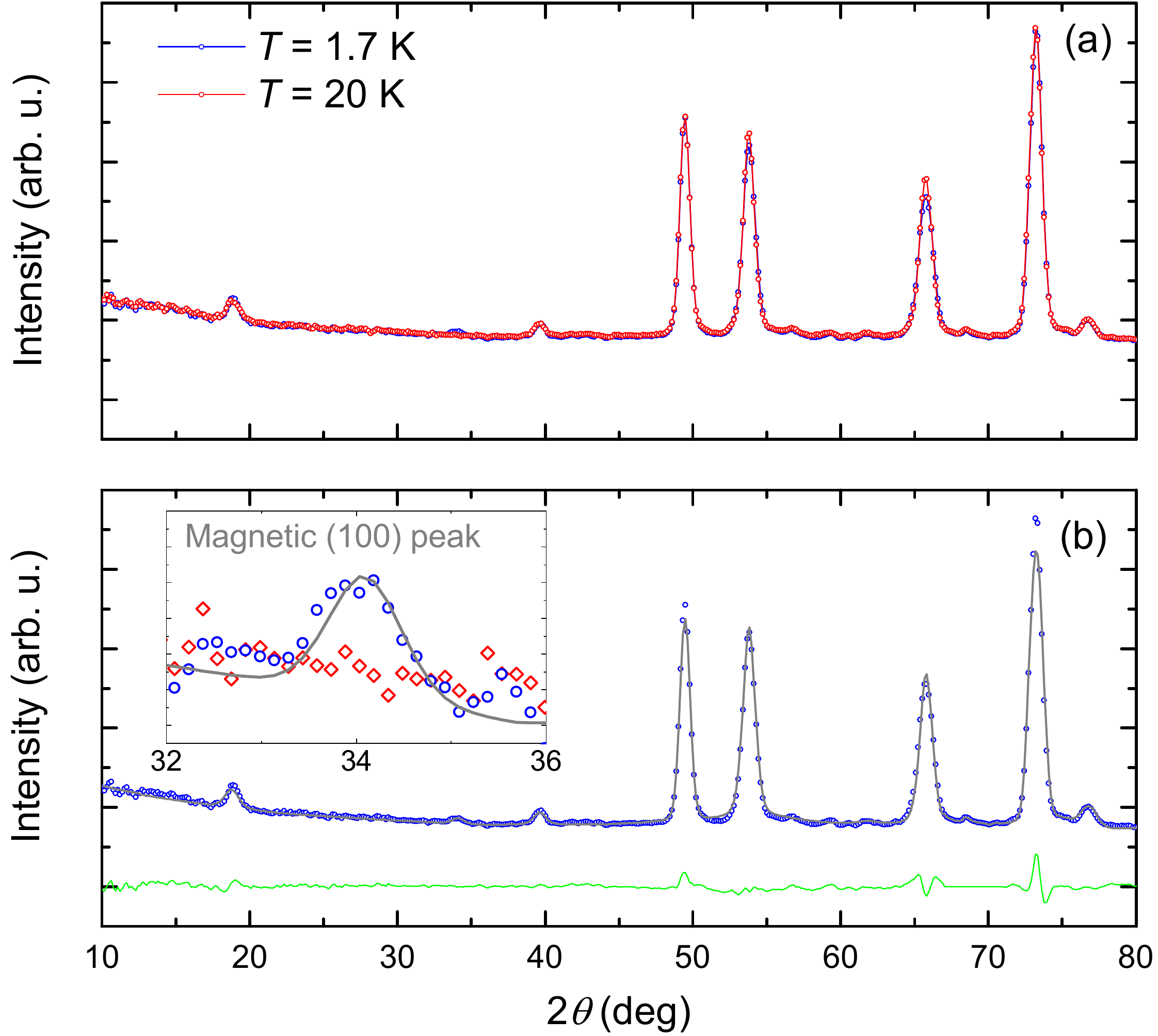}\vspace{-0pt}}
 		 \caption{~(a) Neutron powder diffraction of CeCoSi collected at $T = 1.7$ and 20~K at E6 instrument, HZB facility. (b) Refinement of $T = 1.7$~K diffraction data (blue points - experimental results, gray line - calculated curve, green - difference. Inset shows the zoom of magnetic $(100)$ Bragg peak.)
		}
  	\label{Diffraction}\vspace{-18pt}	
\end{figure}

For our experiment on CeCoSi the LET time-of-flight spectrometer had the special advantage of the so-called multi-repetition mode~\cite{bewley2011let}, which allows one to perform the measurements with several incident neutron energies at the same pulse. Thereby, we could optimize the incident neutron energies in a way to simultaneously measure magnon and CEF excitations, and therefore decrease the counting time needed for a scan at a given temperature and pressure by a factor of two. In our experiment we collected data with three $E_{\mathrm{i}} = 3.43, 6.8$ and 19~meV in the high flux mode~\footnote{Using of $E_{\mathrm{i}} > 19$~meV could provide better energy range for studying the CEF excitations, but would strongly decrease the neutron flux on the sample.}.
To further decrease the background scattering we used a small radial collimator with Gd$_2$O$_3$ painted blades and acceptance diameter of $\sim~4$~mm, which was installed directly on the pressure cell inside the cryostat.
The data were collected in the temperature range 1.7--100~K. The pressure was calculated from the applied press force taking into account the data from CNCS experiment. 

The recorded data were reduced and analyzed using \textsc{JANA2006}~\cite{JANA}, \textsc{DAVE}~\cite{AzuahKneller09}, \textsc{Mantid}~\cite{Mantid} and \textsc{LAMP}~\cite{lamp} software packages. 
\begin{samepage}
Specific-heat measurements were carried out using a commercial PPMS from Quantum Design at temperature range 1.8--300~K.
\end{samepage}

\section{Experimental results}

\subsection{Magnetic structure at ambient pressure}
To characterize the magnetic structure of CeCoSi we measured neutron powder diffraction using the E6 diffractometer at HZB. The powder diffraction patterns were collected  at $T = 1.7$ and 20~K, i.e. below and above the \TN\ and the experimental results are shown in Fig.~\ref{Diffraction}(a). One can see that with decreasing temperature a new weak magnetic satellite appears at $2\theta \approx34~^{\circ}$ (see inset in Fig.~\ref{Diffraction}(b)). The peak can be indexed as $\mathbf{k}=(100)$ (note that the (100) nuclear reflection is forbidden for the $P4/nmm$ space group).

We performed magnetic group representation analysis using JANA2006 software and found that the magnetic symmetry group $Pmm'n$ provide the best fit of our dataset. The low-temperature diffraction pattern along with the calculated curve are shown in Fig.~\ref{Diffraction} (b), and one can see a good agreement ($R_{\mathrm{nuc}} = 2.45~\%$ and $R_{\mathrm{mag}} = 4.84~\%)$. The lattice parameters of the CeCoSi at $T = 1.7$~K were determined to be $a = 3.9967(8)$~\AA\ and $c = 6.937(1)$~\AA\ with the space group $P4/nmm$ (values in brackets denote the $1\sigma$ error of the least-squares fitting throughout the paper). 

The magnetic structure (schematically shown in Fig.~\ref{MagneticStructure}) turned out to be a collinear antiferromagnetic stacking of ferromagnetic Ce layers along the $c$-axis, with the moments pointing along the $[100]$ direction. The ordered Ce moment is as small as $m_{\mathrm{Ce}} = 0.37(6)~\mu_{\mathrm{B}}$. 
It is worth noting that even though our results are consistent with data obtained for the isostructural 
CeCoGe~\cite{chevalier2004reinvestigation}, both analysis are based on a single $(100)$ magnetic reflection, and therefore should be considered with care. Further single-crystal neutron diffraction experiments are highly desirable to confirm the proposed magnetic structure.

\subsection{Spin excitations at ambient pressure}

\subsubsection{Magnon excitations}
\begin{figure}[t]	
\center{\includegraphics[width=1\linewidth]{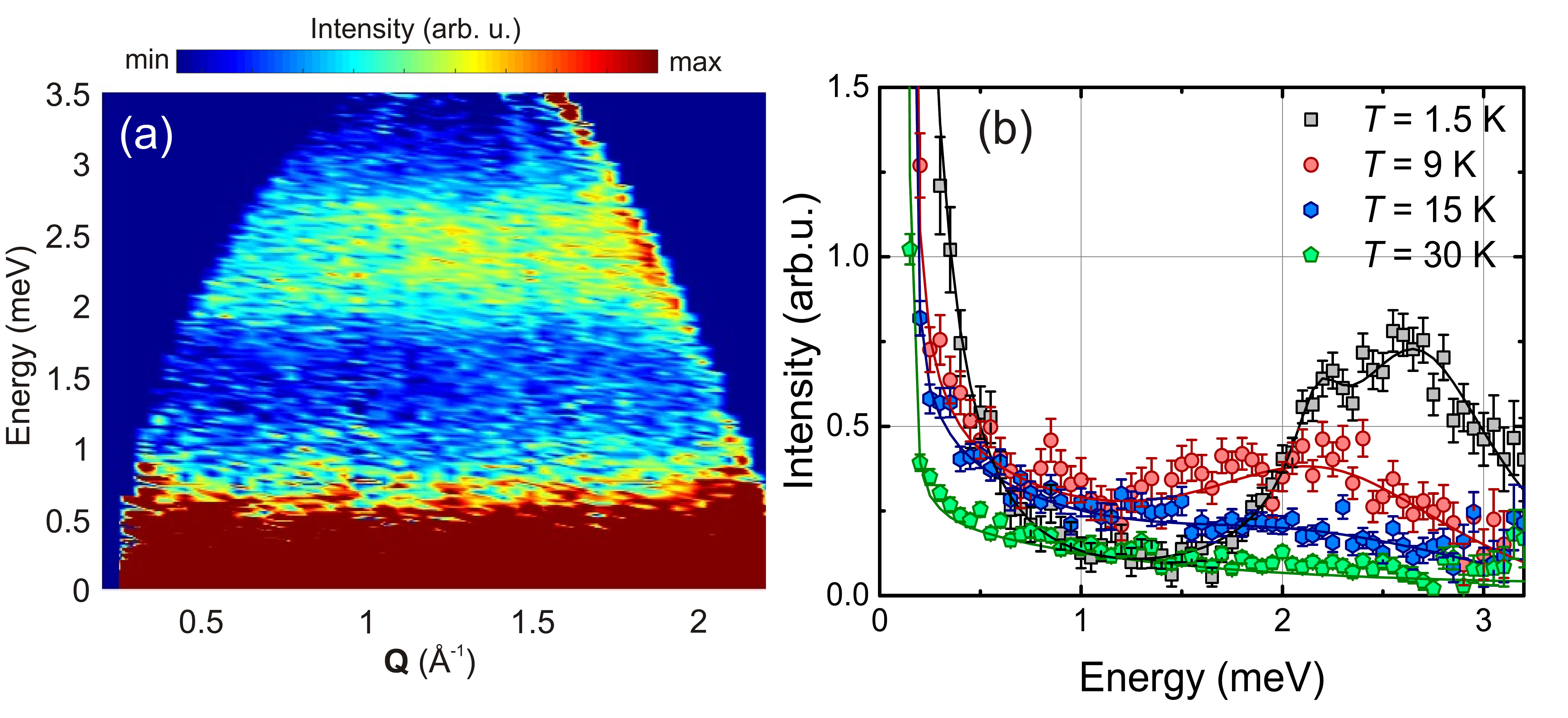}\vspace{-0pt}}
 		 \caption{~(a) ~Low energy INS spectra of CeCoSi measured at IN6 instrument at $T = 1.7$~K with $E_{\mathrm{i}} = 3.86$~meV.
		(b)~Temperature dependence of the energy spectra of CeCoSi integrated within $\mathbf{Q}$ = [1--1.5]~\AA$^{-1}$. Error bars throughout the text represent one standard deviation ($1\sigma$ error).
		}
  	\label{IN6}\vspace{-12pt}	
\end{figure}

To explore the low-energy excitations of CeCoSi we performed powder INS measurements at the spectrometer IN6 at ILL at $T = 1.7 - 100$~K. Figure~\ref{IN6} shows the energy spectra collected with $E_{\mathrm{i}} = 3.86$~meV and integrated within $\mathbf{Q}$ = [1--1.5]~\AA$^{-1}$. The low temperature spectrum consists of a strong gapped magnon band at $E \approx 2.5$~meV. With increasing temperature above \TN\ the gap closes and the spectral weight transfers to the quasielastic channel as expected for a conventional antiferromagnet. 

\vspace{-0pt}

Note that the characteristic energy of the magnetic excitations in CeCoSi is $\sim$~2.5~meV, which is approximately three time higher than the energy associated with the Neel ordering of Ce moments ($T_{\mathrm{N}} = 8.8$~K $\approx 0.75$~meV). This may indicate the presence of magnetic frustration or low-dimensional magnetic behavior of the system. Unfortunately, the powder spectrum appears to be almost featureless, which does not allow us to extract specific details of the underlying magnetic interactions. Therefore, the determination of the low-energy spin Hamiltonian, which should contain at least 3 exchange interaction plus 3 parameters describing the anisotropy of the exchanges, requires further detailed single-crystal INS measurements. 

\subsubsection{CEF excitations} \label{sec:CEF_excitations}
\begin{figure}[b]	
\center{\includegraphics[width=\linewidth]{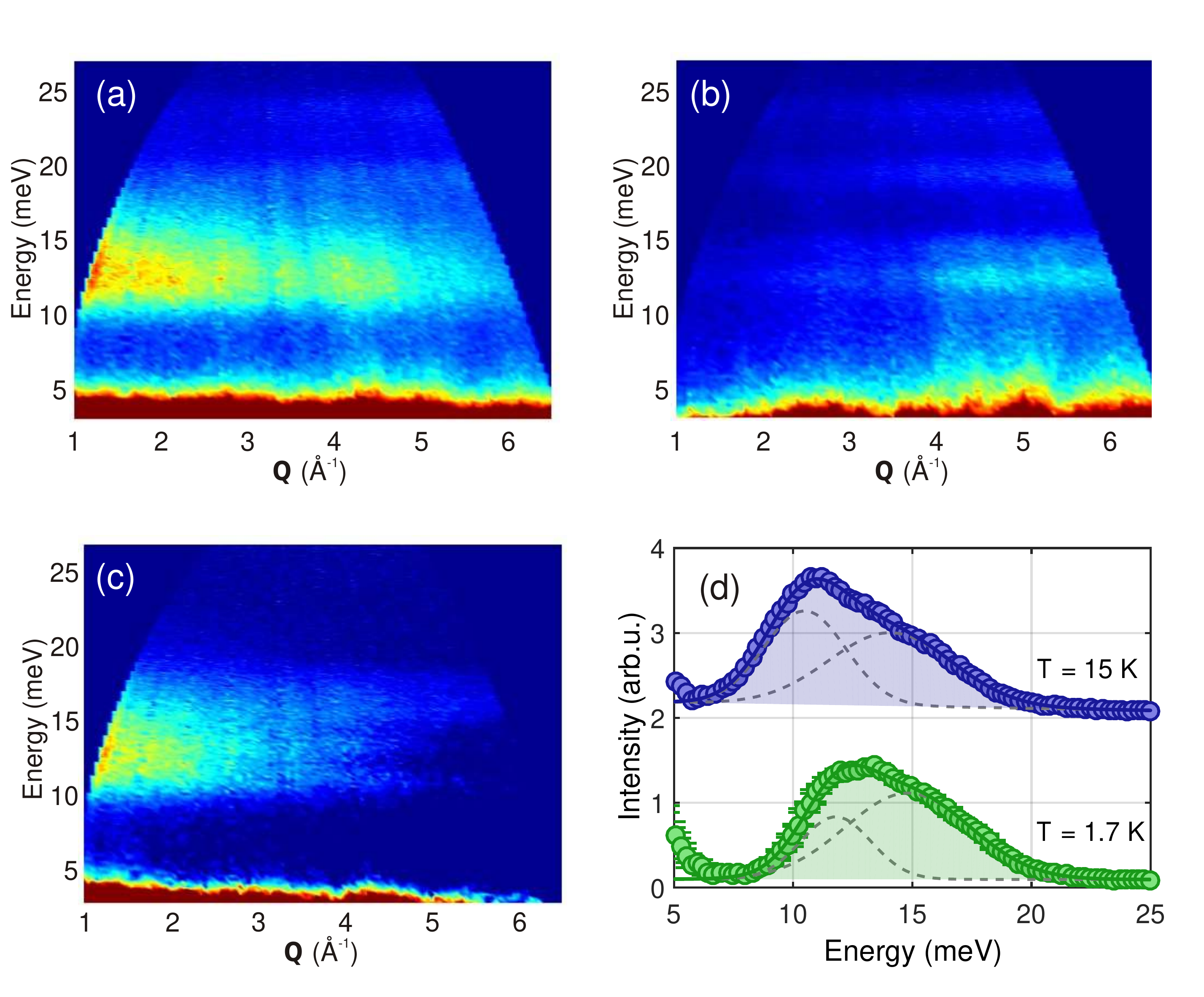}\vspace{-0pt}}
 		 \caption{High-energy INS powder spectra of CeCoSi (a) and LaCoSi (b) measured at the instrument IN4 at $T = 1.7$~K with $E_{\mathrm{i}} = 31.95$~meV.
		(c)~Magnetic signal obtained after subtraction of the scaled LaCoSi spectrum ($\alpha = 1.2$) from the CeCoSi dataset.
		(d)~Background subtracted excitation spectra of CeCoSi above and below the \TN. Grey dotted lines show the deconvolution of the signal into two Gaussian functions. The data are integrated within $\mathbf{Q}$~=~[1--3]~\AA$^{-1}$ and are vertically shifted for clarity.
		}
  	\label{IN4Summary}\vspace{-12pt}	
\end{figure}

Ce$^{3+}$ in CeCoSi has a $J = 5/2$ ground state multiplet, which splits into three doublets under the action of a tetragonal CEF. Thereby, one can expect to observe two CEF transition in an INS spectrum. To characterize the CEF Hamiltonian in CeCoSi we performed INS measurements of CeCoSi and LaCoSi at the TOF instrument IN4 of the Institut Laue-Langevin. 
The spectra of both samples collected at $T = 1.7$~K with $E_{\mathrm{i}} = 31.95$~meV are displayed in Figs.~\ref{IN4Summary}(a, b)~\footnote{We also measured spectra with higher $E_{\mathrm{i}} =  67.6$~meV, but no additional magnetic excitations were observed in the spectra.}. The spectrum of LaCoSi shows strong optical phonon bands, with their intensities increasing with $\mathbf{Q}$ because of the phonon form factor. The spectrum of CeCoSi shows similar phonon bands at large  $\mathbf{Q}$, but in addition exhibits broad magnetic excitations at energies $E \approx 10$--20~meV. 

\vspace{-0pt}

To obtain the magnetic signal of CeCoSi -- $S_{\mathrm{M}}(\mathbf{Q},\hbar\omega)$ we directly subtracted the scaled phonon contribution estimated using the LaCoSi data~\footnote{We also tried a classical approach proposed by A. P. Murani for CeSn$_3$~\cite{murani1983magnetic}, and the results are essentially identical.}. To find the scaling coefficient $\alpha$, we took an energy cut at high momentum, which is dominated by the phonon contribution in both La and Ce samples, because of the phonon and magnetic form-factors. To compensate the difference of the sample masses and scattering lengths we scaled the LaCoSi dataset to get the best agreement between the spectra. Then, we used the obtained coefficient $\alpha$ to scale the LaCoSi spectrum in the whole $\mathbf{Q}$-range and subtract it from the CeCoSi spectrum $S_{\mathrm{M}}(\mathbf{Q},\hbar\omega) = S_{\mathrm{Ce}}(\mathbf{Q},\hbar\omega) - \alpha{}S_{\mathrm{La}}(\mathbf{Q},\hbar\omega)$. The magnetic spectrum after subtraction is displayed in Fig.~\ref{IN4Summary} (c).

To qualitatively extract the positions of CEF peaks we integrated the magnetic spectrum at $\mathbf{Q} =$ [1--3]~\AA$^{-1}$. Two representative curves taken at $T = 1.7$ and 15~K are shown in Fig.~\ref{IN4Summary}(d). Note that the error introduced in the energy cuts when not considering the magnetic form factor and the missing data at small $\mathbf{Q}$ for higher energies, is well below the symbol size of the data points and similar in size to the statistical error.
One can see that the peak shape is rather asymmetric and can not be fitted with a single peak function and therefore, to qualitatively extract the peak positions we fitted the curves with two Gaussian peaks. We found that the peaks are located at $E_1 = 10.49(6)$~meV and $E_2 = 14.1(2)$~meV at $T = 15$~K, i.e. above \TN, and their positions slightly shift in the antiferromagnetic phase at $T = 1.7$~K ($E_1 = 11.78(6)$~meV and $E_2 = 14.8(3)$~meV) due to the splitting of the ground state doublet by an exchange field. It is worth noting that the CEF excitations are broader then the instrumental resolution, which may be due to the interaction with phonons~\cite{vcermak2019magnetoelastic}, hybridization with the conduction band electrons or magnetic dispersion.

\subsection{Specific heat}
To check whether the broad asymmetric peak observed in the INS spectra indeed consists of two CEF excitations we carefully measured the heat capacity of the CeCoSi and LaCoSi samples over a wide temperature range $T = 1.8$--300~K using a PPMS. Specific heat of LaCoSi sample was used as a blank to estimate the phonon contribution and calculate the magnetic contribution in CeCoSi. 

\begin{figure}[t]	
\center{\includegraphics[width=.8\linewidth]{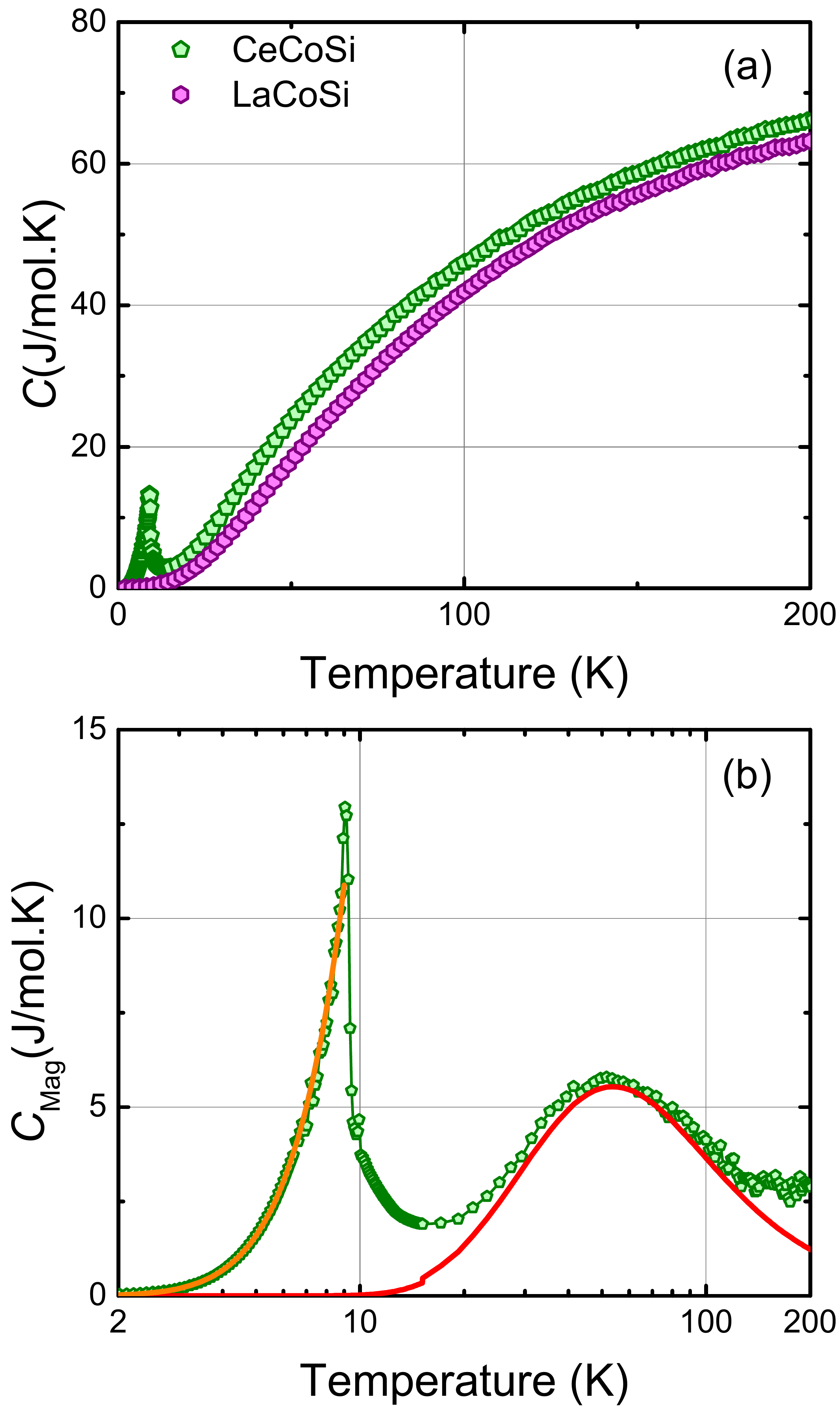}\vspace{-10pt}}
 		 \caption{~(a) Temperature dependences of CeCoSi and LaCoSi specific heat $C(T)$. 
 		  (b)~Magnetic part $C_{\mathrm mag}$ of the specific heat of CeCoSi versus temperature $T$ in a semilogarithmic plot. The solid red and orange lines shows the fits of CEF and magnon contributions to the specific heat using Eq.~(\ref{MultiSchottky}) and Eq.~(\ref{MagnonSpecHeat}) respectively.
		}
  	\label{SpecificHeat}\vspace{-12pt}	
\end{figure}\vspace{-0pt}	
The raw data and the magnetic heat capacity $C_{\mathrm{mag}}$ after subtraction of the phonon contribution are shown in Fig.~\ref{SpecificHeat}. $C_{\mathrm{mag}}(T)$ exhibits two anomalies: a sharp peak at  \TN\ and a broad Schottky-like anomaly with a maximum at $T^* = 51.5$~K. 

First of all, we focus on the high-temperature part of the specific heat curve. One can see that the absolute value of the specific heat $C(T^*) = 5.7$~J/mol$\cdot$K significantly exceeds the 3.65~J/mol$\cdot$K expected for a simple Schottky anomaly for a doublet-doublet transition. In contrast, the $C(T^*)$ is only slightly lower than 6.31~J/mol$\cdot$K -- the peak specific heat expected for a doublet-quartet transition. This indicates that the anomaly is caused by two close standing CEF transitions. 
Also, from the $T^*$ we can estimate the energy gap between the doublet and excited quasi-quartet states $ \Delta \approx 11.8$~meV. Note that this result is in a very good agreement with the mean energy of two doublets observed in our INS measurements $(E_1+E_2)/2 = 12.3$~meV.
To qualitatively calculate the high-temperature magnetic specific heat of CeCoSi we used the standard equation for the specific heat of a discrete $n$-level system:
\begin{eqnarray}
C(T) =  N_{\mathrm{Av}}k_{\mathrm{B}}  \frac{\delta}{\delta T} \Big( \frac{1}{\mathcal{Z}} \sum_{i = 1}^n \ E_{\mathrm{i}}e^{-\frac{E_{\mathrm{i}}}{k_{\mathrm{B}} T} }\Big)
 \label{MultiSchottky}
\end{eqnarray}
where $E_{\mathrm{i}}$ are energies of states and $\mathcal{Z}$ is a partition function. Using Eq.~(\ref{MultiSchottky}) and transition energies $E_1 = 10.49(6)$~meV and $E_2 = 14.09(21)$~meV determined by INS above \TN\ we calculated the magnetic specific heat of CeCoSi, and the results are plotted in Fig.~\ref{SpecificHeat}(b) by the red line. The good agreement between calculated and measured specific heat curves provides another evidence that the CEF transition energies determined by INS are valid. The deviation between the measured and the calculated specific heat curves at high temperature above $\sim 130$~K are caused by the inaccuracy due to subtraction of a massive phononic contribution, which dominates at high temperature.

The low-temperature part of the specific heat contains information about the magnon density-of-state due to the magnetic ordering. For instance, the specific heat of the 3D Heisenberg AFM follows a simple power law $C \propto T^3$ due to the 3D gapless dispersion with $\hbar\omega \propto \mathbf{k}$. On the other hand, if the system has a magnon gap one would expect an activation behavior $C \propto e^{-\Delta/k_\mathrm{B}T}$. 
For the gapped magnons in a three-dimensional magnetic metal the low-temperature part of the specific heat can be expressed as~\cite{continentino2001anisotropic}:
\begin{align}
C(T) = \gamma{}T + b \Delta^{\frac{7}{2}} T^{\frac{1}{2}}e^{-\Delta/k_\mathrm{B}T}\Big(1 + \frac{39}{20}\Big(\frac{T}{\Delta}\Big) + \frac{51}{32}\Big(\frac{T}{\Delta}\Big)^2\Big).
\label{MagnonSpecHeat}
\end{align}
The first term $\gamma{}T$ describes the electronic contribution to the specific heat; $b$ is the constant inversely proportional to the spin-wave velocity $b \propto \Big(\frac{1}{D}\Big)^3$.

We fitted the low-$T$ part of our specific heat curve ($2 < T < \frac{2}{3}$\TN $\approx 6$~K) using Eq.~\eqref{MagnonSpecHeat}.
The fitted curve is shown in Fig.~\ref{SpecificHeat}(b) by orange line, and one can see the perfect agreement between the experimental and calculated curves. It is interesting to note that the extrapolation of our fit function to higher temperature up to 8~K provides surprisingly good description of the observed specific heat data.

The fitted parameters were found to be $\gamma = 23.9(6)$~mJ/mol$\cdot$K$^2$ and $\Delta/k_\mathrm{B} = 12.75(7)$~K. It is worth noting that the gap determined from the specific heat measurements is of the order of the ordering temperature of CeCoSi.

\subsection{Magnetic excitations under hydrostatic pressure}

\begin{figure}[t]	
\center{\includegraphics[width=.9\linewidth]{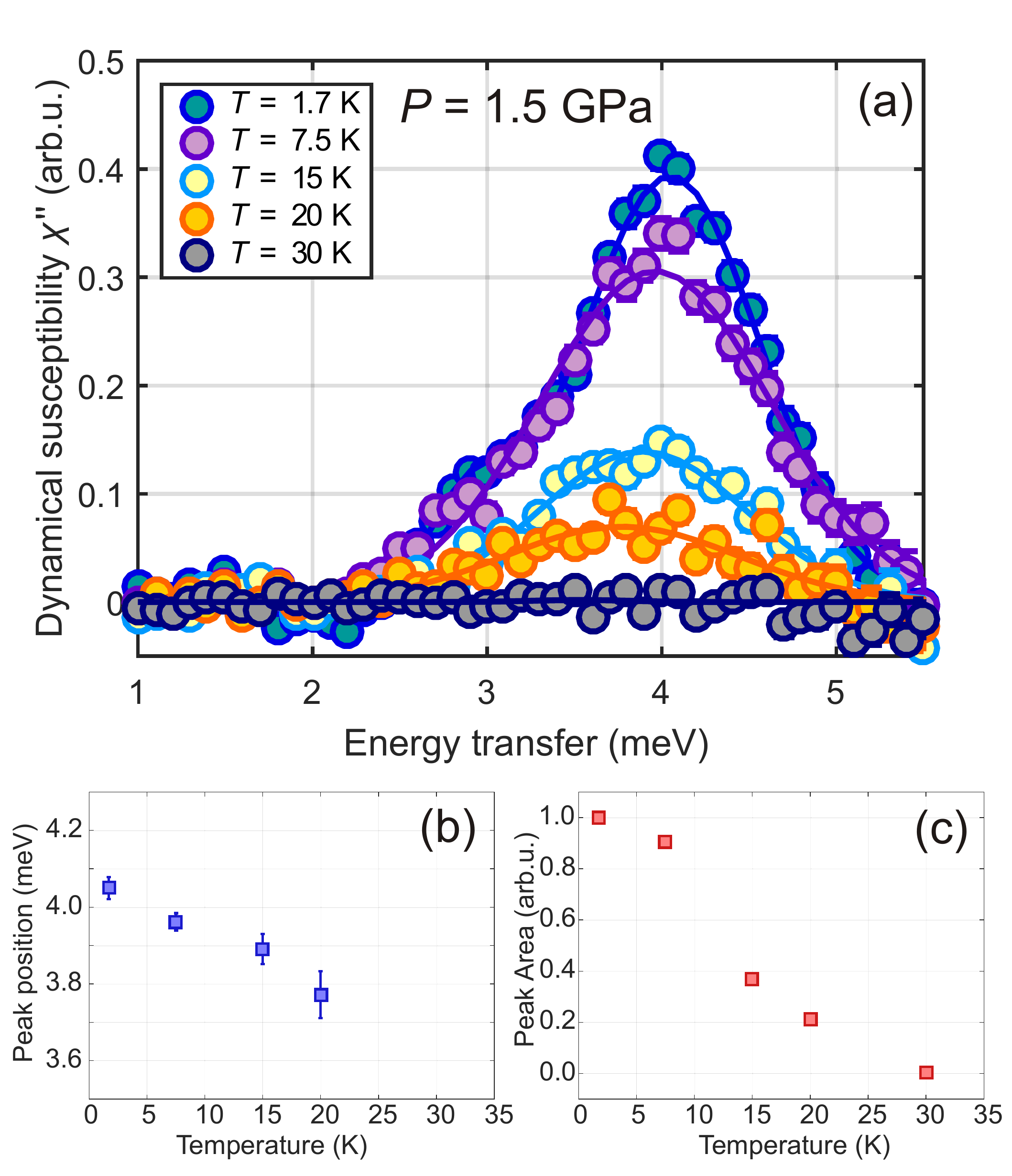}\vspace{-10pt}}
 		 \caption{~Temperature dependence of INS signal at $P = 1.5$~GPa. 
(a) INS spectra taken at  $P = 1.5$~GPa with $E_{\mathrm{i}} = 6.8$~meV and integrated within $\mathbf{Q} = [0.5$--$2.5]$~\AA$^{-1}$. 
(b,c) Magnon peak position and integrated intensity as a functions of temperature. 
		}
  	\label{LETCuts}\vspace{-12pt}	
\end{figure}\vspace{-0pt}

We start our presentation of the pressure-induced evolution of the spin dynamics in CeCoSi with the spectra collected at the LET spectrometer. 
Note that the pressure cell produces a massive background signal. In order to determine the nonmagnetic scattering we used the LaCoSi spectrum measured under similar conditions and the procedure described in Sec.~\ref{sec:CEF_excitations} assuming that $S_{\mathrm{M}}(\mathbf{Q},\hbar\omega) = S_{\mathrm{Ce}}(\mathbf{Q},\hbar\omega) - \alpha\cdot S_{\mathrm{La}}(\mathbf{Q},\hbar\omega)$. However, even without the subtraction a strong broad excitation band at $E \approx 4$~meV is clearly seen in the spectrum (the raw spectra obtained on the LET spectrometer are presented in Appendix~\ref{append}, Fig.~\ref{Raw_LET}(a-d)).

\begin{figure}[b]	
\center{\includegraphics[width=.8\linewidth]{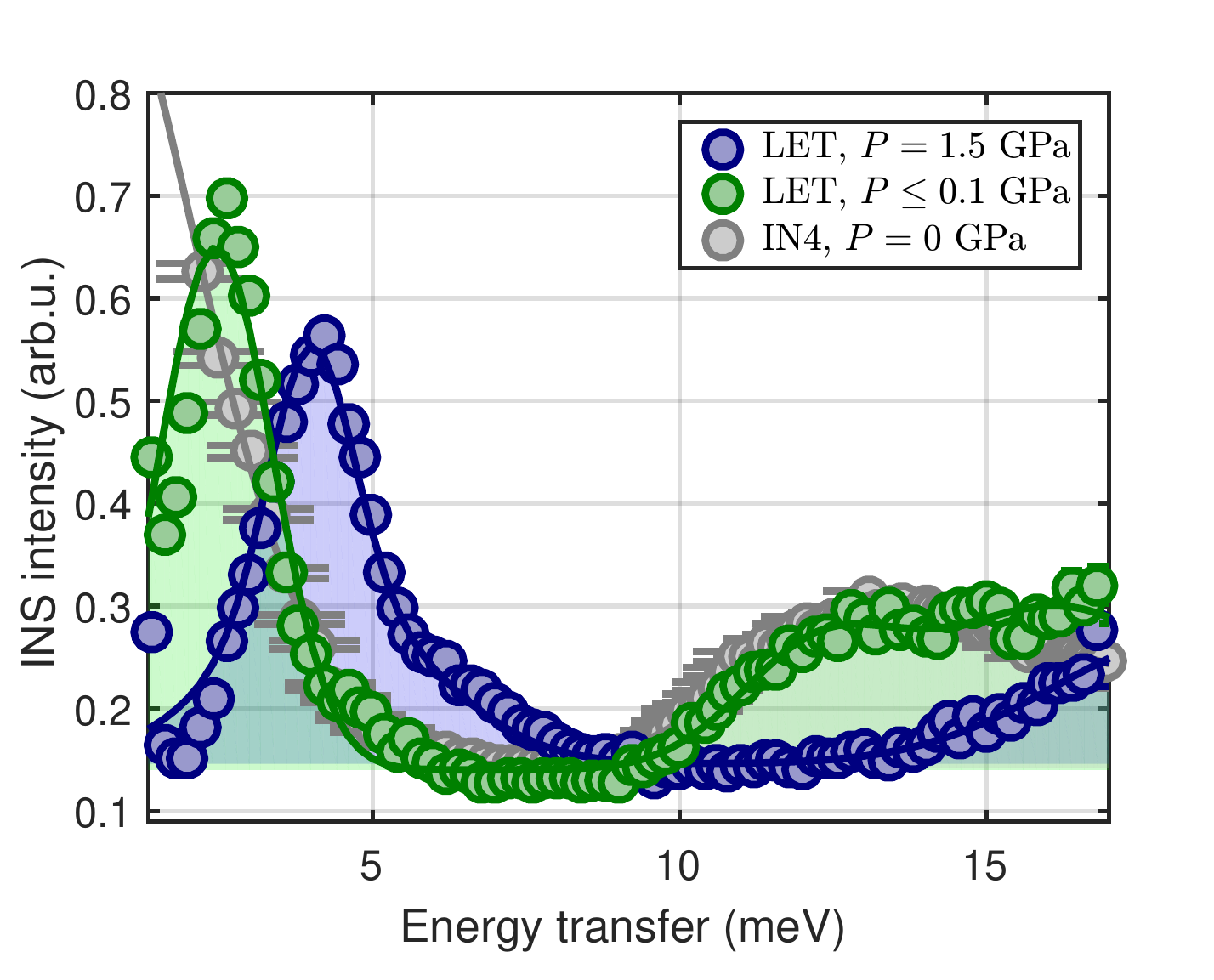}\vspace{-10pt}}
 		 \caption{~Background subtracted INS spectra of CeCoSi taken at $T = 1.7$~K on LET and IN4 instruments.  The data are integrated within $\mathbf{Q}$ = [0.5--2.5]~\AA$^{-1}$ 
		}
  	\label{LETCEF}\vspace{-12pt}	
\end{figure}\vspace{-0pt}

As was discussed above, in this experiment we did not have a pressure sensor in the cell, and the pressure of 1.5~GPa was calculated from the applied press force taking into account $\sim 10$~\% loss, while cooling down to 1.7~K, which results in the relatively large estimated uncertainty of the pressure determination of $\sim 0.25$~GPa. For this reason we decided to study the $T$ dependence of the observed mode at fixed $P$. We subtracted the background and Bose-corrected all obtained spectra measured with $E_{\mathrm{i}} = 6.8$~meV. The resulting $\chi''(\hbar\omega)$ curves integrated within $\mathbf{Q} = [0.5$--$2.5]$~\AA$^{-1}$ are shown in Fig.~\ref{LETCuts} (a). Increasing temperature induces a decrease of the mode intensity, and slightly shifts down the peak position. Fits of these parameters are presented in Fig.~\ref{LETCuts}(b, c) and one can see that the magnon mode intensity disappears below the detection limit at $T = 30$~K. This result is in a reasonable agreement with the phase diagram of CeCoSi, which shows transition temperature of $\sim 35$~K at $P \approx 1.5$~GPa. 

In order to check the consistency of our results with the zero pressure data we also measured the spectra at almost ambient condition ($P \leq 0.1$~GPa) at 1.7~K. The resulting spectrum along with the 1.5~GPa data and results of the IN4 experiment are shown in Fig.~\ref{LETCEF}. The position of the CEF excitations obtained in the LET experiment perfectly coincides with the IN4 results indicating that we can reliably extract information about both CEF and magnon excitations from the LET data. It is interesting to note that the pressure of $P = 1.5$~GPa significantly shifts or suppresses the intensity of the CEF excitations as clearly seen in Fig.~\ref{LETCEF}. 

\begin{figure}[t]	
\center{\includegraphics[width=1\linewidth]{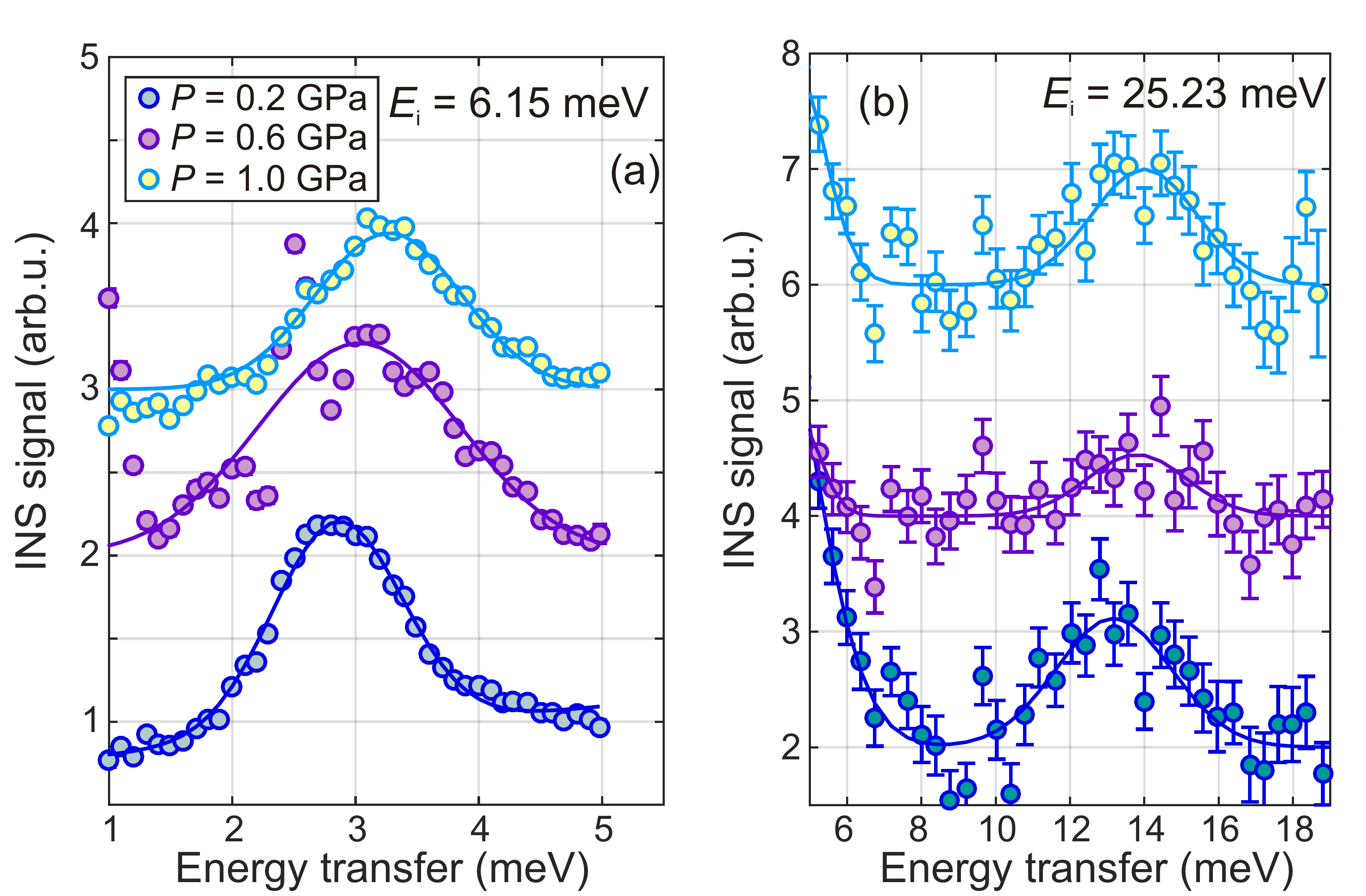}\vspace{-10pt}}
 		 \caption{~Magnetic signal of CeCoSi after background subtraction measured using CNCS at $T = 1.7$~K with $E_{\mathrm{i}} = 6.15$~meV (a) and $E_{\mathrm{i}} = 25.23$~meV (b). The data are integrated within $\mathbf{Q} = [1$--$ 2.5]$~\AA$^{-1}$ and $\mathbf{Q} = [1.8$--$ 3]$~\AA$^{-1}$ in left and right panels, respectively. Different data sets are shifted vertically for ease of viewing. 
		}
  	\label{CNCS}\vspace{-12pt}	
\end{figure}\vspace{-0pt}

We now focus on the pressure dependence in more detail and present data obtained on the CNCS spectrometer. Figure~\ref{CNCS} shows the summary of the background subtracted signal at 0.2, 0.6 and 1.0~GPa collected with two incident neutron energies~\footnote{Note that the signal at 0.6~GPa was measured during a separate beamtime with slightly different setup, which did not allowed us to subtract the background precisely. This problem causes sharp features in Fig.~\ref{CNCS}(a) and is responsible for the lower intensity of 0.6~GPa signal shown in the Fig.~\ref{CNCS}(b).}. The $E_{\mathrm{i}} = 6.15$~meV spectra presented in the left panel display a rather strong magnon peak, which position gradually shifts upon increasing pressure. It is worth noting that already at the lowest pressure of 0.2~GPa the position of the peak is slightly higher than the one obtained in our IN6 experiment at ambient pressure. 

The high-energy data have much stronger background due to the phonon scattering from the pressure cell. The data after subtracting the background contribution are shown in panel (b) of Fig.~\ref{CNCS}. At $P = 0.2$~GPa we found a weak peak at an energy of $\sim13$~meV. Its position is close to $E_1 = 11.78(6)$~meV and $E_2 = 14.81(26)$~meV observed in the IN4 experiment at ambient pressure. The position of the peak also shits to higher energies with pressure. However, the signal-to-noise ratio is much worse in the 25.23~meV dataset compared to the 6.15~meV one, as can be seen from the ratio between the neutron count rate and errorbars in the two panels of Fig.~\ref{CNCS}, and the 13~meV peak has an intensity, which exceeds the background level by 2-4 standard deviations only. Taking into account that the positions and intensities of the peaks would depend on the details of the subtraction procedure, we would like to point out that the obtained result should be considered with a reasonable caution, because we can not unambiguously proof a magnetic origin of the observed peak, which can be just an artefact of the background subtraction procedure~\footnote{Investigation of the form-factor of the excitation is not possible due to the cell phonon scattering at high $\mathbf{Q}$ above $\sim$3~\AA$^{-1}$.}.

On the other hand, the fitting of the low-energy peak (Fig.~\ref{CNCS}~(a)) is rather robust and self-consistent, independent on subtraction details. Accordingly, we can conclude that the energy of the magnon mode indeed increases with the pressure, whereas the observation of the CEF excitations and their $P$ dependence is much more questionable.

\section{Discussion}
Our experimental work on CeCoSi has a dual aim: (i) to characterize the magnetic ground state and excitations of CeCoSi at ambient pressure using a combination of different techniques and (ii) to study how the magnetic excitations evolve with pressure. Analyzing the results of neutron powder diffraction in the AFM phase and at ambient pressure we detected only one weak magnetic satellite peak, which appears below \TN\ and can be indexed as the $(100)$ reflection. This result is consistent with a simple commensurate antiferromagnetic structure, previously proposed for isostructural CeCoGe~\cite{chevalier2004reinvestigation}. The Ce moments are aligned along the $[100]$ direction and carry an ordered moment of only $m_{\mathrm{Ce}} = 0.37(6)~\mu_{\mathrm{B}}$, which is significantly reduced compared to the moment of free Ce$^{3+}$ with $m_{\mathrm{Ce}} = 2.14~\mu_{\mathrm{B}}$. However, in Ce systems the CEF is comparatively strong and therefore the $J = 5/2$ multiplet is split in such a way that the energy of the first excited CEF level is in general much larger than \TN. Then, the size of the ordered moment is limited to that of the CEF ground state doublet, which for the easy CEF direction is in the range 1 -- 2.5~$\mu_{\mathrm{B}}$~\cite{fischer1990mean}. But even compared to the lower bound, the observed value is small. 
This is striking, because the $4f$ entropy collected just above \TN\ is close to $R$ln2. 
This indicates that only a very small amount of the $4f$ entropy connected with the ground state CEF doublet is collected above \TN. Since the onset of correlations is always associated with the reduction of entropy, the amount of correlation above \TN\ within the CEF ground state doublet cannot be large, and therefore these correlations should not be able to result in a strong reduction of the ordered moment far below \TN.
Thus, the standard scenarios invoked to account for a reduced size of ordered moments, the presence of Kondo interaction or frustration, does not apply since in there scenario the reduction of the moment in the ordered state far below \TN\ is connected with a reduction of the entropy collected at \TN~\cite{besnus1992correlation, bredl1978specific}.
Another alternative, which is presently discussed for a number of ferromagnetic systems, the ordering along the hard CEF direction, where the available CEF moment can be very small~\cite{kruger2014fluctuation, hafner2019kondo}, seems to be unlikely because susceptibility data do not indicate a strong anisotropy~\cite{tanida2019successive}. Thus, the origin of the strong reduction of the ordered moment is a further mystery in this system.

Both the inelastic neutron scattering and the specific heat results demonstrate that the first excited and the second excited CEF doublets are close-by in energy and at a mean energy of the order of 12~meV (about 140~K). This corroborates the doubts on the possibility of a quadrupolar transition in the temperature range 10 -- 40~K expressed in the introduction. Because Ce$^{3+}$ is a Kramers ion and thus each CEF doublet does not bear a quadrupolar degree of freedom, a quadrupolar transition has to be an induced one, a process which is well-known for magnetic order in singlet systems. However that requires the ordering temperature to be larger than typically $\Delta/2$ where $\Delta$ is the energy splitting between the involved CEF ground state and excited state. Thus in the present case $T_{\mathrm{Q}}$ should be larger than about 70~K. According to established results for the magnetic singlet-singlet case, in such induced ordering processes the ratio between the ordering temperature $T_{\mathrm{c}}$ and $\Delta$ is given by $T_{\mathrm{c}}/\Delta = \mathrm{artanh}(\Delta/J)$ where $J$ is a coupling parameter~\cite{Cooper1972Magnetic}. The artanh function results in an extreme fast, almost vertical drop of $T_{\mathrm{c}}$ with decreasing $J$ for $T_{\mathrm{c}}/\Delta < 0.5$, making the realization and stabilization of such a low $T_{\mathrm{c}}/\Delta < 0.5$ very difficult and very unlikely. This almost vertical dependence of $T_{\mathrm{c}}/\Delta$ is e.g. incompatible with the very smooth an almost linear increase of the proposed quadrupolar transition in CeCoSi under pressure shown in~\cite{tanida2018substitution}. This analysis confirms that the transition at $T = 12$~K at $P = 0$~\cite{tanida2019successive} and at $T \approx 36$~K at $P = 1.5$~GPa are connected with an unconventional order. However, it is worth noting that in contrast to the results of~\cite{tanida2019successive} we did not observe any indication of phase transition at 12~K at ambient pressure in our data.

In the low-energy spectrum we observed magnon excitations with a characteristic energy of $E^* \approx 2.5$~meV (29~K) at ambient pressure. It is worth noting that the excitation energy scale of the magnons exceeds by more than three times the ordering temperature of \TN\ = 8.8~K. One possible explanation is a quasi-2D magnetic structure of the material (see Fig.~\ref{MagneticStructure}) with much stronger exchange interactions within the $ab$-planes and only weak coupling along the $c$ direction ($J_{\mathrm{c}} <J_{\mathrm{ab}}$). In that case, short-range fluctuations within the $ab$ plane will survive at temperatures above the \TN. Indeed, we were able to resolve a broad paramagnon inelastic peak at 9 and 10~K, inline with such a scenario, whereas at higher temperatures all spectral weight is transferred to the quasielastic channel. 

However, the properties of the CEF ground state deduced from our analysis raise a further problem (see appendix~\ref{Appendix:B}). The wave function of this CEF ground state corresponds to a $c$-axis moment of $m_c = 0.97~\mu_{\mathrm{B}}$ and a basal plane moment of $m_a = 0.56~\mu_{\mathrm{B}}$, thus it is not very anisotropic. This weak anisotropy of the local moment cannot account for the large gap in the magnetic excitation spectrum of the ordered state deduced from the INS and specific heat results. Furthermore the very weak anisotropy observed in the magnetic susceptibility indicates that the anisotropy of the magnetic interactions is also weak. Thus this large gap in the magnetic excitations is a further open problem in this unusual system.

\begin{figure}[t]	
\center{\includegraphics[width=.9\linewidth]{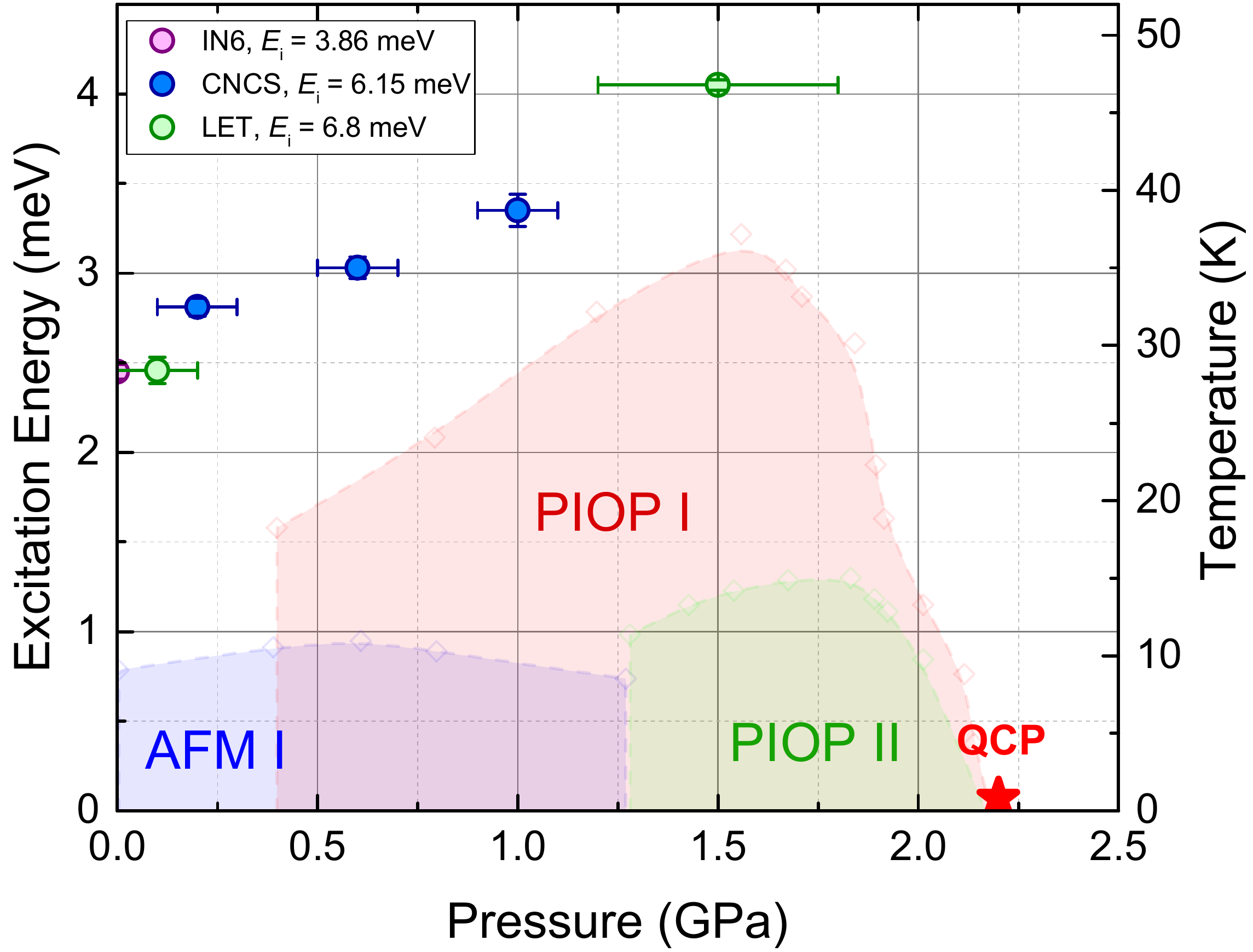}\vspace{-10pt}}
 		 \caption{~Temperature-Pressure phase diagram of CeCoSi from Ref.~\cite{lengyel2013temperature} and the $P$-dependence of the magnon mode (solid circles) at $T = 1.7$~K. Shaded areas represent different phases -- AFM and two pressure-induced ordered phases (PIOP). Note that the right and left axis are shown on the same scale $E = k_{\mathrm{B}}T$.
		}
  	\label{PhaseDiagram}\vspace{-12pt}	
\end{figure}\vspace{-0pt}

Summarizing the results of the low-energy INS experiments under pressure, we found that the energy of the magnon mode gradually evolves from 2.5~meV at ambient pressure to $\sim 4$~meV at 1.5~GPa (see Fig.~\ref{PhaseDiagram}). At this pressure the energy scale of the magnetic excitations is comparable to the transition temperature of the pressure-induced phase (4~meV $\approx$ 46~K). Note that these results are not in favor of the metaorbital transition scenario, because the last implies a sharp, abrupt change of the ordered moment and the magnon excitation energy as a consequence, which is in contrast to the gradual pressure-induced evolution observed in our measurements. 

The pressure dependence of the CEF excitations is less clear: in the low-pressure (0.2~GPa) CNCS experiment we found a weak peak, close to the positions of the CEF excitations observed at zero-pressure measurements. The peak position changes only slightly with pressures up to 1~GPa. On the other hand the results of the LET experiment unambiguously showed that at 1.5~GPa the CEF levels move out of their original location. One possible explanation in much worse signal-to-noise ratio in the high-energy CNCS measurements, which cast some doubts on the CNCS results. However, if one looks at the signature in the resistivity, the high-$T$ ordering is something new which appears quite abruptly at $P \geq 1.4$~GPa, while the observed effects at lower pressure were different and order of magnitude weaker. Therefore, $P_1 = 1.4$~GPa was explicitly introduced to highlight this strong change in~\cite{lengyel2013temperature}. Thus the appearance of the strong anomaly in $\rho(T)$ at $P \geq P_1$ may be related to the dramatic change of CEF excitations between 1 and 1.5~GPa, indicating that there is a real strong difference between the orderings above and below the $P_1$, as was suggested in~\cite{lengyel2013temperature}. A single crystal neutron diffraction under pressure is should be performed to resolve this question and clarify the order parameter of the PIOP.

\section{Conclusion}

To summarize, we performed a comprehensive experimental investigation of CeCoSi by means of neutron scattering and specific heat measurements. At ambient pressure CeCoSi orders into a simple AFM structure with a surprisingly small ordered moment of only $m_{\mathrm{Ce}} = 0.37(6)~\mu_{\mathrm{B}}$ and exhibits spin excitations on two different energy scales: low-energy collective magnons at $\sim$ 2.5~meV and two CEF transitions at $\sim 12$~meV. 
The application of hydrostatic pressure up to 1.5~GPa causes a gradual shift of the magnon band towards higher energies and significantly modifies the CEF splitting scheme at 1.5~GPa.
The obtained results are not in favor of the metaorbital scanario~\cite{hattori2010meta}, which was proposed to describe the origin of the pressure-induced phases in CeCoSi~\cite{lengyel2013temperature}.

\section*{ACKNOWLEDGMENTS}
We acknowledge A. S. Sukhanov for stimulating discussions, C. Goodway for help with the high-pressure experiments at ISIS Neutron and Muon source and K. A. Nikitina for assistance with the CEF analysis. 
This research used resources at the Spallation Neutron Source, a DOE Office of Science User Facility operated by Oak Ridge National Laboratory.
S.E.N. acknowledges support from the International Max Planck Research School for Chemistry and Physics of Quantum Materials (IMPRS-CPQM).
D.G.F and C.G. aknowlege support form the German Research Fundation (DFG) through grants GE~602/4-1 and Fermi-NEst.

\newpage

\appendix

\section{Raw INS data}\label{append}

\begin{figure}[h]	
\center{\includegraphics[width=1\linewidth]{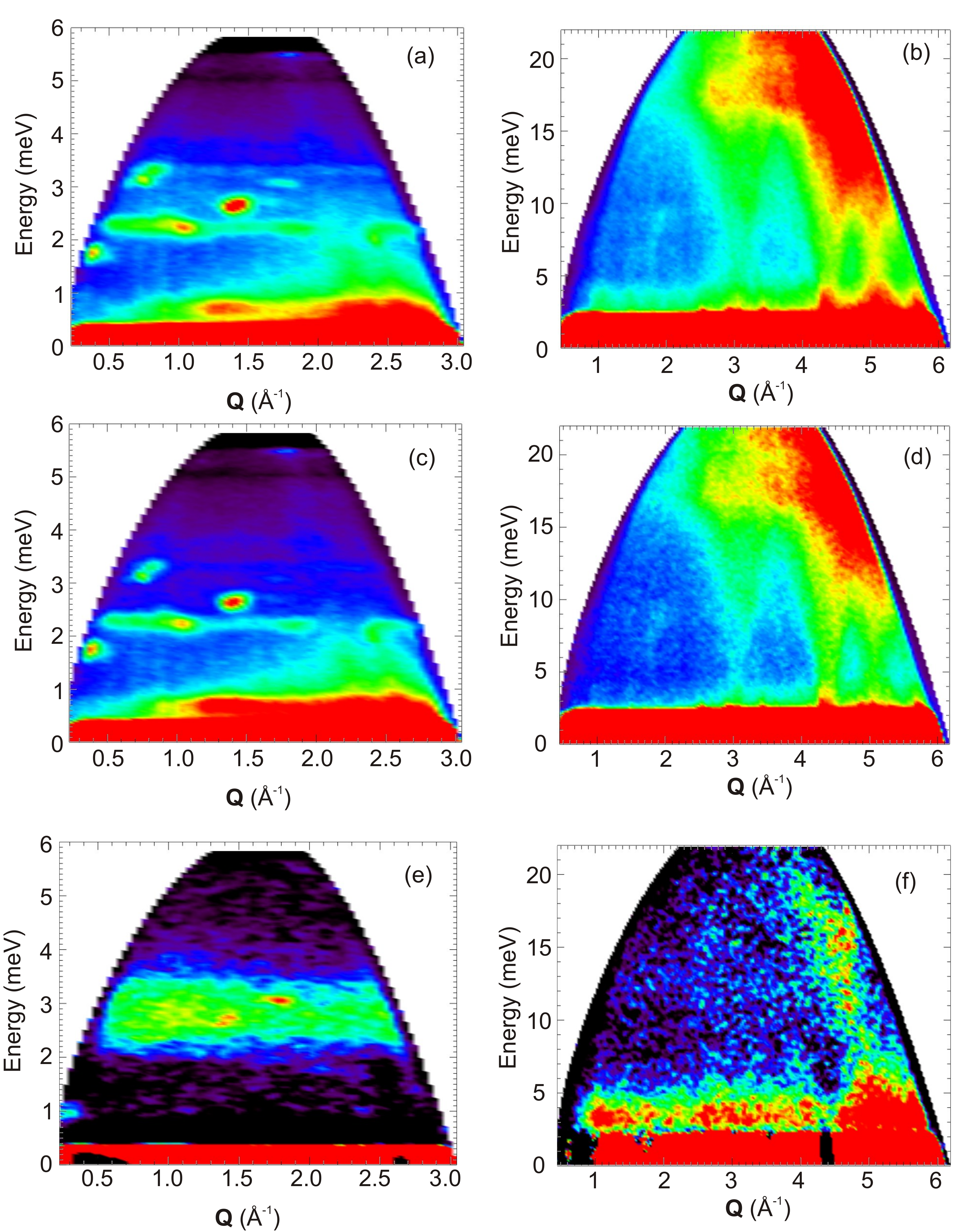}\vspace{-10pt}}
 		 \caption{~INS spectra of CeCoSi (a, b) and LaCoSi (c, d) measured at $T = 1.7$~K on the CNCS spectrometer with $E_{\mathrm{i}} = 6.15$~meV (left) and $E_{\mathrm{i}} = 25.23$~meV (right) using a NiCrAl pressure cell at a pressure $P~=~0.2$~GPa.
 		 (e, f)~INS spectra of CeCoSi after the subtraction of the nonmagnetic LaCoSi contribution. The intensities were scaled by $\times5$ with respect to the raw spectra (a-d) to highlight the observed excitations.}
  	\label{Raw_CNCS}\vspace{-12pt}	
\end{figure}

\begin{figure}[t]	
\center{\includegraphics[width=1\linewidth]{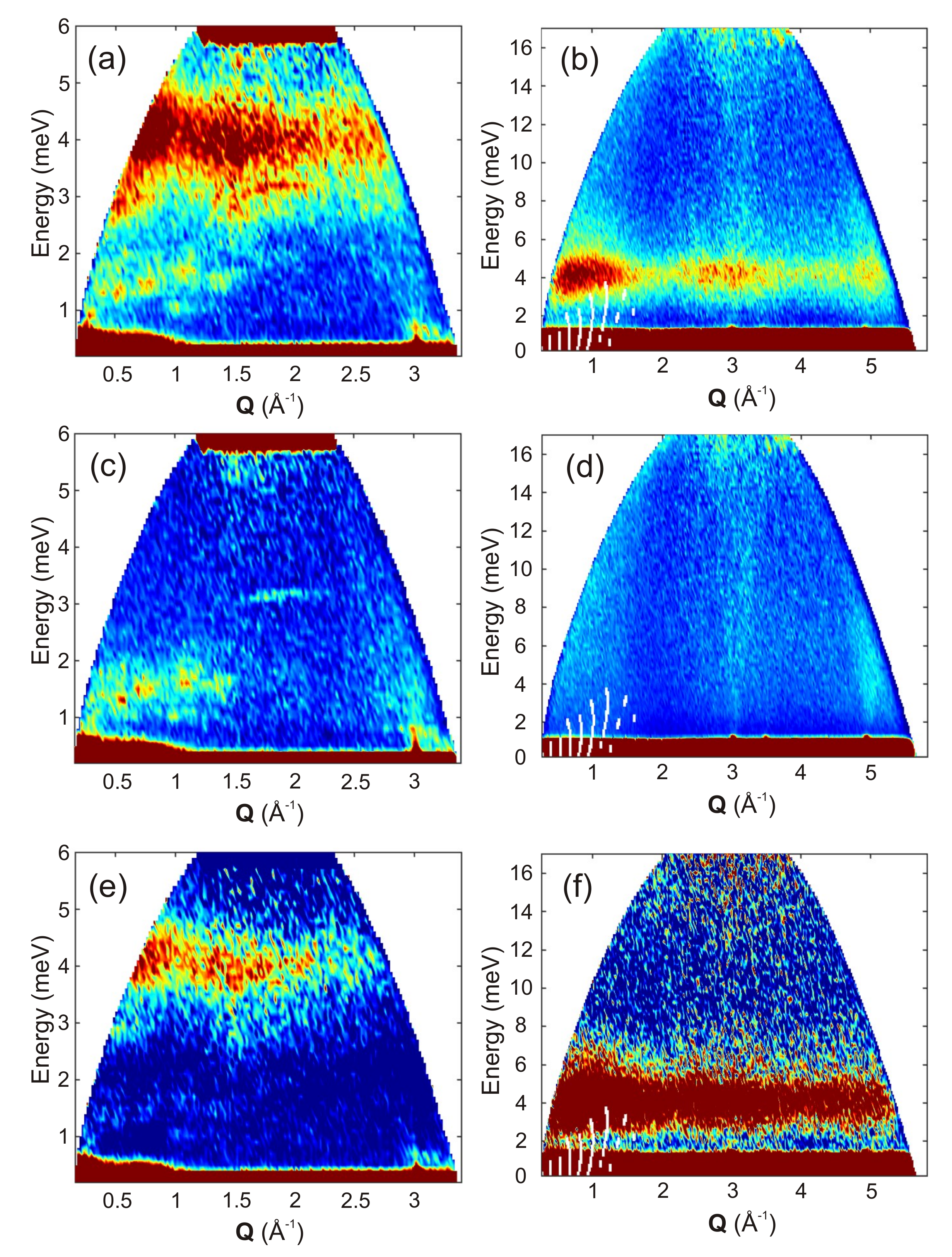}\vspace{-10pt}}
 		 \caption{~INS spectra of CeCoSi (a, b) and LaCoSi (c, d) measured at $T = 1.7$~K on the LET spectrometer with $E_{\mathrm{i}} = 6.8$~meV (left) and $E_{\mathrm{i}} = 19$~meV (right) using a NiCrAl pressure cell at a pressure $P~=~1.5$~GPa.
 		 (e, f)~INS spectra of CeCoSi after the subtraction of the nonmagnetic LaCoSi contribution. The intensity of the (f) panel was scaled by $\times5$ with respect to (b, d) panels to highlight the excitations.}
  	\label{Raw_LET}\vspace{-12pt}	
\end{figure}

Figures~\ref{Raw_CNCS} and \ref{Raw_LET} show the raw $S(\mathbf{Q},\omega)$ of CeCoSi and LaCoSi samples and the magnetic spectra $S_{\mathrm{M}}(\mathbf{Q},\omega)$ after phonon/background subtraction measured on CNCS and LET spectrometers, respectively.

\vspace{-0pt}

\vspace{-0pt}

\section{Analysis of CEF Hamiltonian}\label{Appendix:B}

The Ce ions in CeCoSi occupy a position with $4mm$ point symmetry, and thereby the crystalline electric field Hamiltonian of the Ce$^{3+}$ ion will include only three $B_l^m$ coefficients. In Stevens notation it can be written as:
\begin{align}
	\mathcal{H} = B_2^0 O_2^0 + B_4^0 O_4^0 + B_4^4 O_4^4
	\label{Eq:CEF}
\end{align}

In the paramagnetic phase the Hamiltonian~\eqref{Eq:CEF} exhibits three Kramers doublet, therefore at low temperatures $\Delta_{\mathrm{CEF}} \gg k_{\mathrm{B}}T$ the INS spectrum consists of two transition. In section~\ref{sec:CEF_excitations} we report the observation of two CEF levels at $T = 15$~K with energies $\Delta_1 = 10.49(6)$~meV and $\Delta_2 = 14.1(2)$~meV. The relative intensity ratio is $I_1/I_2 \approx 0.83$.
Taking into account these results along with the temperature dependence of the magnetic susceptibility reported on a single crystalline sample~\cite{tanida2019successive} we performed a fitting of the Hamiltonian~\eqref{Eq:CEF}.

As the first step we fitted the transition energies. For that we defined the deviation as:
\begin{align}
 \chi =	\Big(\sum_i^2 (E_i^{\mathrm{calc}} - E_i^{\mathrm{obs}})^2\Big)^{\frac{1}{2}},
	\label{Eq:deviation}
\end{align}
and made a ``brute-force'' search of the $B_l^m$ coefficients within the parameter space of $B_2^0 = [-2, 2]$~meV; $B_4^0 = [-1, 1]$~meV; $B_4^4 = [-1, 1]$~meV
with a step size of 2~$\mu$eV.

\begin{figure}[t]	
\center{\includegraphics[width=1\linewidth]{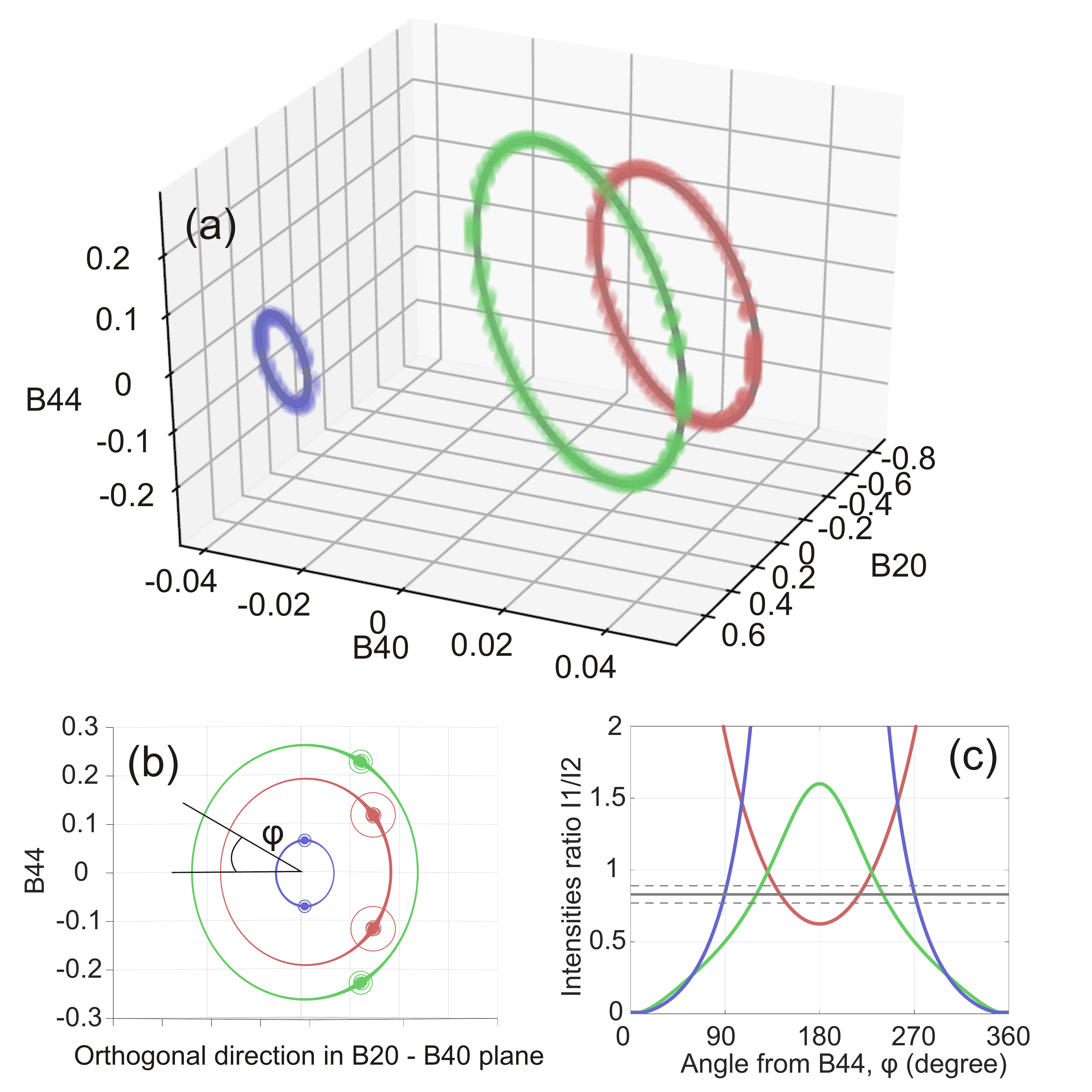}\vspace{-10pt}}
 		 \caption{~Fitting of the CEF Hamiltonian. 
 		 (a)~Semi-transparent color rings show the parameter space of the Hamiltonian~\eqref{Eq:CEF} with low cost function Eq.~\eqref{Eq:deviation} $\chi < 0.2$~meV. The three colors are used to indicate different parameter regions. The gray solid lines show the individual fitting of those regions using SVD as described in the main text.
 		 (b)~The same regions plotted in the parameter plane, which is roughly orthogonal to the rings. The size of the points indicates the deviation from the experimental $I_1/I_2$ ratio defined as: Size = $\frac{1}{|I_1/I_2(\mathrm{Calc}) - I_1/I_2(\mathrm{Obs})|}$. The
 		 sketch explains the definition of the $\varphi$ angle. 
 		 (c)~$I_1/I_2$ as a function of $\varphi$ for each parameter region. Solid and dotted gray lines represent the observed ratio $I_1/I_2 = 0.83(6)$.
		}
  	\label{Fig:CEF1}\vspace{-12pt}	
\end{figure}\vspace{-0pt}

To find the $E^{\mathrm{Obs}}$ we calculated the eigenvalues of the Hamiltonian~\eqref{Eq:CEF} with the given set of $B_l^m$. Figure~\ref{Fig:CEF1}(a) shows the sets of parameters which satisfy the condition of $\chi < 0.2$~meV in three dimensional $B_l^m$ space. 

One can see that the solutions form three rings. We applied density-based spatial clustering DBSCAN as implemented in sklearn library~\cite{scikit} to separate them. 
Then, we described each of the obtained rings using a singular value decomposition approach (SVD). 
In simple, for each ring we shifted the center of the coordinate system to the mean value: $\hat{B}_l^m \rightarrow B_l^m + \overline{B_l^m}$. After that, we searched for a matrix $M$, which would rotate/deform the coordinate system in a way to approximate the dataset by a unit circle. 
Thus, we presume that the solutions have an elliptical shape, which seems to be valid with the experimental precision of the energy determination. 
Figure~\ref{Fig:CEF1}(a) shows the eigenvalues calculated by this ``brute-force'' method and the fitting with SVD. One can see the excellent agreement between the curves.
The obtained values for $\overline{B_l^m}$ and the rotation matrices $M$ for each solution rings are given in table~\ref{tab:SVD} and can be used to recalculate each solution ring.

As the next step we calculated the ratio of the transition intensities $I_1/I_2$ for the obtained rings using the standard equation for the INS transition intensity:
\begin{align}
	I(\psi_i \rightarrow \psi_j) \propto 
	\sum_{\alpha = x, y, z}   
	|\langle \psi_j |J_{\alpha}| \psi_i \rangle|^2.
	\label{Eq:intensity}
\end{align}
Figure~\ref{Fig:CEF1}(c) shows $I_1/I_2$ for all rings as a function of the angle $\varphi$ from the $B_4^4$ axis, for definition of $\varphi$ see Fig.~\ref{Fig:CEF1}(b). Based on $I_1/I_2 = 0.83(6)$ as determined in the experiment we considered the solutions within the interval $I_1/I_2 = [0.77$--$0.89]$. 
In that case we obtained six main regions of the parameter space (two for each ring), which meet the required ratio $I_1/I_2$. Note that both INS spectra and the magnetization depend only on absolute values, but not on the sign of the $B_4^4$ coefficient, therefore we could further reduce the number of the considered regions to three.

\begin{figure}[t]	
\center{\includegraphics[width=1\linewidth]{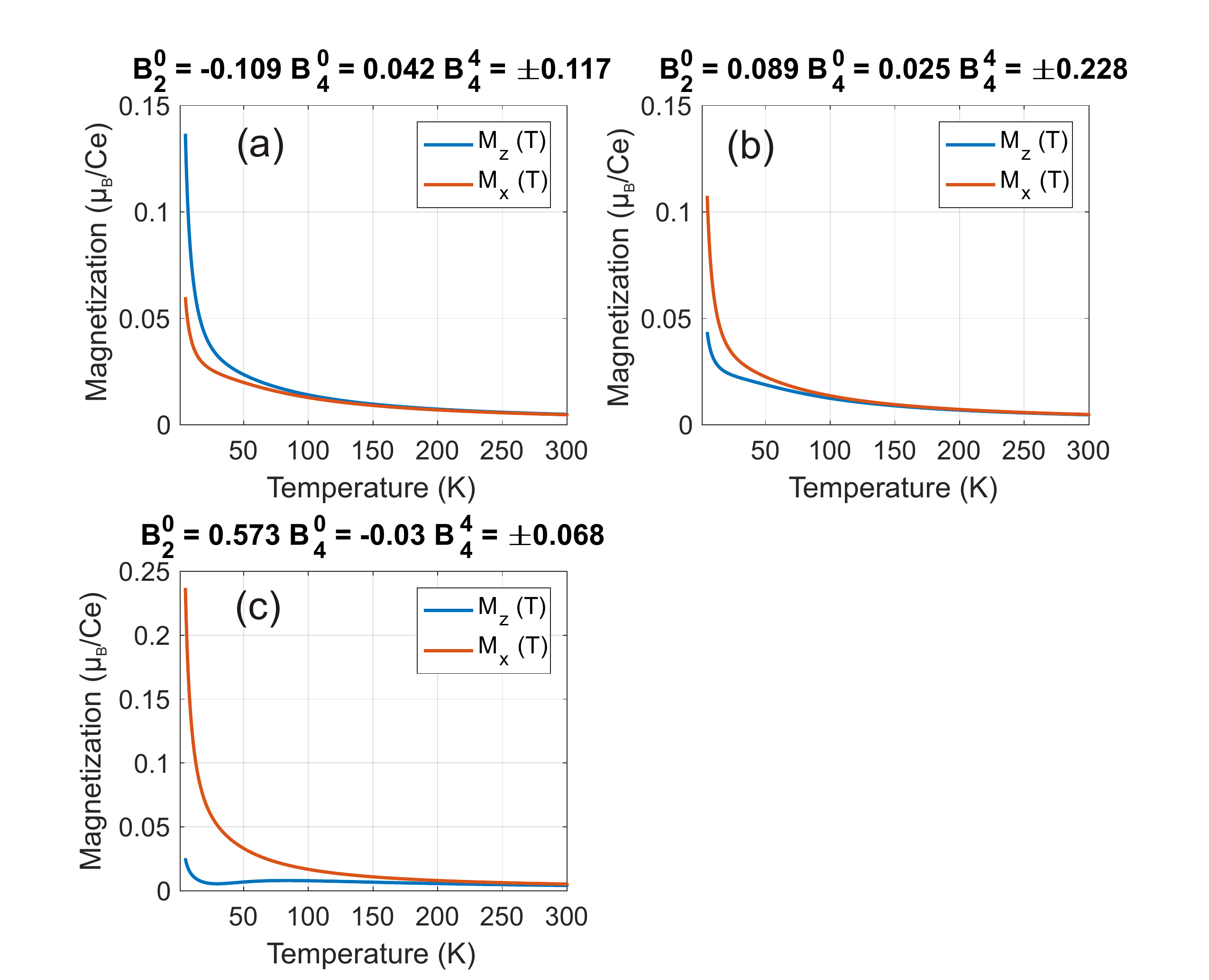}\vspace{-10pt}}
 		 \caption{~Calculated temperature dependence of the magnetization for the Hamiltonian~\eqref{Eq:CEF}. The magnetization curves were obtained for a magnetic field of $B = 1$~T applied along the $a$ and $c$ axes and three sets of $B_l^m$ parameters as indicated in each panel. $B_l^m$ parameters are given in meV units.
		}
  	\label{Fig:CEF2}\vspace{-12pt}	
\end{figure}

We calculated the magnetic susceptibilities along the $c$ direction and along the basal plane corresponding to the central value from each of those areas, and compared the result of these calculations with the experimental susceptibilities reported in ~\cite{tanida2019successive}. Note that all solutions from the same parameter area result into quantitatively similar $\chi(T)$ curves. 
Figure~\ref{Fig:CEF2} shows the calculated temperature dependence of the magnetization. One can see that the magnetization shown in panel (c), which corresponds to the blue ring in Fig.~\ref{Fig:CEF1}~(a), exhibits a very strong easy-plane anisotropy due to the large positive $B_2^0$ coefficient. This result is in clear disagreement with the reported susceptibility data.
On the other hand, two other sets of $B_l^m$ imply a relatively isotropic susceptibility, at least down to 15--20~K. In the experiment, there was a small, but noticeable hierarchy $M_c > M_a$, therefore we believe that the first solution shown in Fig.~\ref{Fig:CEF2} (a) $B_2^0 = -0.109$~meV; $B_4^0 = 0.042$~meV; $B_4^4 = \pm 0.117$~meV provides the best fit of all experimental data.

The transition energies, wavefunctions and symmetry representations of the doublets for $B_2^0 = -0.109$~meV; $B_4^0 = 0.042$~meV; $B_4^4 = 0.117$~meV  are given below: 
\begin{align}
E_0 &= 0;  \hspace{0.3cm}       &\psi_{0\pm} &= \mp 0.306| \pm\frac{5}{2} \rangle \pm 0.95 | \mp\frac{3}{2}\rangle; \hspace{0.3cm} &\Gamma_7^{(1)}, \nonumber\\
E_1 &= 10.78~\mathrm{meV}; \hspace{0.3cm}    &\psi_{1\pm} &=  0.95| \pm\frac{5}{2} \rangle + 0.306 | \mp\frac{3}{2}\rangle; \hspace{0.3cm} &\Gamma_7^{(2)}, \nonumber\\
E_2 &= 14.26~\mathrm{meV}; \hspace{0.3cm}     &\psi_{2\pm} &= 1| \pm\frac{1}{2} \rangle; \hspace{0.3cm} &\Gamma_6.\nonumber 
\end{align}

Using these wavefunctions we calculated the magnetic moments of the ground state doublet along $c$ and $a$ directions as: 
\begin{align}
m_{\{\alpha = x, z\}} = g \langle \psi_{0} | J_{\alpha} | \psi_{0}\rangle,
\end{align}
where $g=6/7$ for the Ce$^{3+}$ ion. The moments were found to be $m_x = 0.558~\mu_{\mathrm{B}}$ and $m_z = 0.965~\mu_{\mathrm{B}}$.

\vspace{-0pt}

 \begin{table}[b]
 \caption{~Parameters of SVD fitting for the three rings from Fig.~\ref{Fig:CEF1}(a) (all $\overline{B_l^m}$ in meV)} \label{tab:SVD}
 \begin{ruledtabular}
 \begin{tabular}{p{0.8cm}|p{1cm} | p{1cm} | p{.4cm} | p{3.5cm}}
           & $\overline{B_2^0}$  &$\overline{B_4^0}$& $\overline{B_4^4}$& Transformation matrix  \vspace{0.05cm}\\
           \hline
 Red       & -0.41      & 0.022     & 0               & $\begin{pmatrix}
 -0.3804   & 0        & -0.0001     \\
 -0.0255   & 0        & 0.0009      \\
 0         & 0.1922   & 0  
 \end{pmatrix}$\\
 \hline
 Green      & -0.161     & 0.008    & 0                & $\begin{pmatrix}
 -0.5048 & 0        & -0.0001 \\
 -0.0337 & 0        & 0.001 \\
 0       & 0.2623   &0   
 \end{pmatrix}$\\    
 \hline
 Blue       & 0.574     & -0.03      & 0                & $\begin{pmatrix}
 -0.1295 & 0     & 0 \\
 -0.0087 & 0     & 0.0005 \\
 0       & 0.068 &0   
 \end{pmatrix}$\\

 \end{tabular}
 \end{ruledtabular}
 \end{table}

\newpage

\bibliography{main}

\begin{thebibliography}{41}%
\makeatletter
\providecommand \@ifxundefined [1]{%
 \@ifx{#1\undefined}
}%
\providecommand \@ifnum [1]{%
 \ifnum #1\expandafter \@firstoftwo
 \else \expandafter \@secondoftwo
 \fi
}%
\providecommand \@ifx [1]{%
 \ifx #1\expandafter \@firstoftwo
 \else \expandafter \@secondoftwo
 \fi
}%
\providecommand \natexlab [1]{#1}%
\providecommand \enquote  [1]{``#1''}%
\providecommand \bibnamefont  [1]{#1}%
\providecommand \bibfnamefont [1]{#1}%
\providecommand \citenamefont [1]{#1}%
\providecommand \href@noop [0]{\@secondoftwo}%
\providecommand \href [0]{\begingroup \@sanitize@url \@href}%
\providecommand \@href[1]{\@@startlink{#1}\@@href}%
\providecommand \@@href[1]{\endgroup#1\@@endlink}%
\providecommand \@sanitize@url [0]{\catcode `\\12\catcode `\$12\catcode
  `\&12\catcode `\#12\catcode `\^12\catcode `\_12\catcode `\%12\relax}%
\providecommand \@@startlink[1]{}%
\providecommand \@@endlink[0]{}%
\providecommand \url  [0]{\begingroup\@sanitize@url \@url }%
\providecommand \@url [1]{\endgroup\@href {#1}{\urlprefix }}%
\providecommand \urlprefix  [0]{URL }%
\providecommand \Eprint [0]{\href }%
\providecommand \doibase [0]{http://dx.doi.org/}%
\providecommand \selectlanguage [0]{\@gobble}%
\providecommand \bibinfo  [0]{\@secondoftwo}%
\providecommand \bibfield  [0]{\@secondoftwo}%
\providecommand \translation [1]{[#1]}%
\providecommand \BibitemOpen [0]{}%
\providecommand \bibitemStop [0]{}%
\providecommand \bibitemNoStop [0]{.\EOS\space}%
\providecommand \EOS [0]{\spacefactor3000\relax}%
\providecommand \BibitemShut  [1]{\csname bibitem#1\endcsname}%
\let\auto@bib@innerbib\@empty
\bibitem [{\citenamefont {Stockert}\ and\ \citenamefont
  {Steglich}(2011)}]{stockert2011unconventional}%
  \BibitemOpen
  \bibfield  {author} {\bibinfo {author} {\bibfnamefont {O.}~\bibnamefont
  {Stockert}}\ and\ \bibinfo {author} {\bibfnamefont {F.}~\bibnamefont
  {Steglich}},\ }\bibfield  {title} {\enquote {\bibinfo {title}
  {{Unconventional quantum criticality in heavy-fermion compounds}},}\ }\href
  {\doibase 10.1146/annurev-conmatphys-062910-140546} {\bibfield  {journal}
  {\bibinfo  {journal} {Annu. Rev. Condens. Matter Phys.}\ }\textbf {\bibinfo
  {volume} {2}},\ \bibinfo {pages} {79--99} (\bibinfo {year}
  {2011})}\BibitemShut {NoStop}%
\bibitem [{\citenamefont {Gegenwart}\ \emph {et~al.}(2008)\citenamefont
  {Gegenwart}, \citenamefont {Si},\ and\ \citenamefont
  {Steglich}}]{gegenwart2008quantum}%
  \BibitemOpen
  \bibfield  {author} {\bibinfo {author} {\bibfnamefont {P.}~\bibnamefont
  {Gegenwart}}, \bibinfo {author} {\bibfnamefont {Q.}~\bibnamefont {Si}}, \
  and\ \bibinfo {author} {\bibfnamefont {F.}~\bibnamefont {Steglich}},\
  }\bibfield  {title} {\enquote {\bibinfo {title} {Quantum criticality in
  heavy-fermion metals},}\ }\href {\doibase 10.1038/nphys892} {\bibfield
  {journal} {\bibinfo  {journal} {Nature Phys.}\ }\textbf {\bibinfo {volume}
  {4}},\ \bibinfo {pages} {186--197} (\bibinfo {year} {2008})}\BibitemShut
  {NoStop}%
\bibitem [{\citenamefont {Si}\ and\ \citenamefont
  {Steglich}(2010)}]{si2010heavy}%
  \BibitemOpen
  \bibfield  {author} {\bibinfo {author} {\bibfnamefont {Q.}~\bibnamefont
  {Si}}\ and\ \bibinfo {author} {\bibfnamefont {F.}~\bibnamefont {Steglich}},\
  }\bibfield  {title} {\enquote {\bibinfo {title} {Heavy fermions and quantum
  phase transitions},}\ }\href {\doibase 10.1126/science.1191195} {\bibfield
  {journal} {\bibinfo  {journal} {Science}\ }\textbf {\bibinfo {volume}
  {329}},\ \bibinfo {pages} {1161--1166} (\bibinfo {year} {2010})}\BibitemShut
  {NoStop}%
\bibitem [{\citenamefont {Stockert}\ \emph {et~al.}(2011)\citenamefont
  {Stockert}, \citenamefont {Arndt}, \citenamefont {Faulhaber}, \citenamefont
  {Geibel}, \citenamefont {Jeevan}, \citenamefont {Kirchner}, \citenamefont
  {Loewenhaupt}, \citenamefont {Schmalzl}, \citenamefont {Schmidt},
  \citenamefont {Si},\ and\ \citenamefont
  {Steglich}}]{stockert2011magnetically}%
  \BibitemOpen
  \bibfield  {author} {\bibinfo {author} {\bibfnamefont {O.}~\bibnamefont
  {Stockert}}, \bibinfo {author} {\bibfnamefont {J.}~\bibnamefont {Arndt}},
  \bibinfo {author} {\bibfnamefont {E.}~\bibnamefont {Faulhaber}}, \bibinfo
  {author} {\bibfnamefont {C.}~\bibnamefont {Geibel}}, \bibinfo {author}
  {\bibfnamefont {H.~S.}\ \bibnamefont {Jeevan}}, \bibinfo {author}
  {\bibfnamefont {S.}~\bibnamefont {Kirchner}}, \bibinfo {author}
  {\bibfnamefont {M.}~\bibnamefont {Loewenhaupt}}, \bibinfo {author}
  {\bibfnamefont {K.}~\bibnamefont {Schmalzl}}, \bibinfo {author}
  {\bibfnamefont {W.}~\bibnamefont {Schmidt}}, \bibinfo {author} {\bibfnamefont
  {Q.}~\bibnamefont {Si}}, \ and\ \bibinfo {author} {\bibfnamefont
  {F.}~\bibnamefont {Steglich}},\ }\bibfield  {title} {\enquote {\bibinfo
  {title} {{Magnetically driven superconductivity in CeCu$_2$Si$_2$}},}\ }\href
  {\doibase 10.1038/nphys1852} {\bibfield  {journal} {\bibinfo  {journal}
  {Nature Phys.}\ }\textbf {\bibinfo {volume} {7}},\ \bibinfo {pages}
  {119--124} (\bibinfo {year} {2011})}\BibitemShut {NoStop}%
\bibitem [{\citenamefont {Kitaoka}\ \emph {et~al.}(1995)\citenamefont
  {Kitaoka}, \citenamefont {Tou}, \citenamefont {q.~Zheng}, \citenamefont
  {Ishida}, \citenamefont {Asayama}, \citenamefont {Kobayashi}, \citenamefont
  {Kohda}, \citenamefont {Takeshita}, \citenamefont {Amaya}, \citenamefont
  {Onuki}, \citenamefont {Geibel}, \citenamefont {Schank},\ and\ \citenamefont
  {Steglich}}]{kitaoka1995nmr}%
  \BibitemOpen
  \bibfield  {author} {\bibinfo {author} {\bibfnamefont {Y.}~\bibnamefont
  {Kitaoka}}, \bibinfo {author} {\bibfnamefont {H.}~\bibnamefont {Tou}},
  \bibinfo {author} {\bibfnamefont {G.}~\bibnamefont {q.~Zheng}}, \bibinfo
  {author} {\bibfnamefont {K.}~\bibnamefont {Ishida}}, \bibinfo {author}
  {\bibfnamefont {K.}~\bibnamefont {Asayama}}, \bibinfo {author} {\bibfnamefont
  {T.~C.}\ \bibnamefont {Kobayashi}}, \bibinfo {author} {\bibfnamefont
  {A.}~\bibnamefont {Kohda}}, \bibinfo {author} {\bibfnamefont
  {N.}~\bibnamefont {Takeshita}}, \bibinfo {author} {\bibfnamefont
  {K.}~\bibnamefont {Amaya}}, \bibinfo {author} {\bibfnamefont
  {Y.}~\bibnamefont {Onuki}}, \bibinfo {author} {\bibfnamefont
  {C.}~\bibnamefont {Geibel}}, \bibinfo {author} {\bibfnamefont
  {C}~\bibnamefont {Schank}}, \ and\ \bibinfo {author} {\bibfnamefont
  {F.}~\bibnamefont {Steglich}},\ }\bibfield  {title} {\enquote {\bibinfo
  {title} {{NMR study of strongly correlated electron systems}},}\ }\href
  {\doibase 10.1016/0921-4526(94)00365-3} {\bibfield  {journal} {\bibinfo
  {journal} {Physica B}\ }\textbf {\bibinfo {volume} {206}},\ \bibinfo {pages}
  {55--61} (\bibinfo {year} {1995})}\BibitemShut {NoStop}%
\bibitem [{\citenamefont {Lengyel}\ \emph {et~al.}(2011)\citenamefont
  {Lengyel}, \citenamefont {Nicklas}, \citenamefont {Jeevan}, \citenamefont
  {Geibel},\ and\ \citenamefont {F.~Steglich}}]{lengyel2011pressure}%
  \BibitemOpen
  \bibfield  {author} {\bibinfo {author} {\bibfnamefont {E.}~\bibnamefont
  {Lengyel}}, \bibinfo {author} {\bibfnamefont {M.}~\bibnamefont {Nicklas}},
  \bibinfo {author} {\bibfnamefont {H.~S.}\ \bibnamefont {Jeevan}}, \bibinfo
  {author} {\bibfnamefont {C.}~\bibnamefont {Geibel}}, \ and\ \bibinfo {author}
  {\bibfnamefont {F}~\bibnamefont {F.~Steglich}},\ }\bibfield  {title}
  {\enquote {\bibinfo {title} {{Pressure tuning of the interplay of magnetism
  and superconductivity in CeCu$_2$Si$_2$}},}\ }\href {\doibase
  10.1103/PhysRevLett.107.057001} {\bibfield  {journal} {\bibinfo  {journal}
  {Phys. Rev. Lett.}\ }\textbf {\bibinfo {volume} {107}},\ \bibinfo {pages}
  {057001} (\bibinfo {year} {2011})}\BibitemShut {NoStop}%
\bibitem [{\citenamefont {Coleman}(2015)}]{coleman2015heavy}%
  \BibitemOpen
  \bibfield  {author} {\bibinfo {author} {\bibfnamefont {P.}~\bibnamefont
  {Coleman}},\ }\bibfield  {title} {\enquote {\bibinfo {title} {{Heavy fermions
  and the Kondo lattice: a 21st century perspective}},}\ }\href
  {https://arxiv.org/abs/1509.05769} {\bibfield  {journal} {\bibinfo  {journal}
  {arXiv:1509.05769}\ } (\bibinfo {year} {2015})}\BibitemShut {NoStop}%
\bibitem [{\citenamefont {Shen}\ \emph {et~al.}(2020)\citenamefont {Shen},
  \citenamefont {Zhang}, \citenamefont {Komijani}, \citenamefont {Nicklas},
  \citenamefont {Borth}, \citenamefont {Wang}, \citenamefont {Chen},
  \citenamefont {Nie}, \citenamefont {Li}, \citenamefont {Lu}, \citenamefont
  {Lee}, \citenamefont {Smidman}, \citenamefont {Steglich}, \citenamefont
  {Coleman},\ and\ \citenamefont {Yuan}}]{shen2020strange}%
  \BibitemOpen
  \bibfield  {author} {\bibinfo {author} {\bibfnamefont {Bin}\ \bibnamefont
  {Shen}}, \bibinfo {author} {\bibfnamefont {Yongjun}\ \bibnamefont {Zhang}},
  \bibinfo {author} {\bibfnamefont {Yashar}\ \bibnamefont {Komijani}}, \bibinfo
  {author} {\bibfnamefont {Michael}\ \bibnamefont {Nicklas}}, \bibinfo {author}
  {\bibfnamefont {Robert}\ \bibnamefont {Borth}}, \bibinfo {author}
  {\bibfnamefont {An}~\bibnamefont {Wang}}, \bibinfo {author} {\bibfnamefont
  {Ye}~\bibnamefont {Chen}}, \bibinfo {author} {\bibfnamefont {Zhiyong}\
  \bibnamefont {Nie}}, \bibinfo {author} {\bibfnamefont {Rui}\ \bibnamefont
  {Li}}, \bibinfo {author} {\bibfnamefont {Xin}\ \bibnamefont {Lu}}, \bibinfo
  {author} {\bibfnamefont {Hanoh}\ \bibnamefont {Lee}}, \bibinfo {author}
  {\bibfnamefont {Michael}\ \bibnamefont {Smidman}}, \bibinfo {author}
  {\bibfnamefont {Frank}\ \bibnamefont {Steglich}}, \bibinfo {author}
  {\bibfnamefont {Piers}\ \bibnamefont {Coleman}}, \ and\ \bibinfo {author}
  {\bibfnamefont {Huiqiu}\ \bibnamefont {Yuan}},\ }\bibfield  {title} {\enquote
  {\bibinfo {title} {{Strange-metal behaviour in a pure ferromagnetic Kondo
  lattice}},}\ }\href {\doibase 10.1038/s41586-020-2052-z} {\bibfield
  {journal} {\bibinfo  {journal} {Nature}\ }\textbf {\bibinfo {volume} {579}},\
  \bibinfo {pages} {51--55} (\bibinfo {year} {2020})}\BibitemShut {NoStop}%
\bibitem [{\citenamefont {Lengyel}\ \emph {et~al.}(2013)\citenamefont
  {Lengyel}, \citenamefont {Nicklas}, \citenamefont {Caroca-Canales},\ and\
  \citenamefont {Geibel}}]{lengyel2013temperature}%
  \BibitemOpen
  \bibfield  {author} {\bibinfo {author} {\bibfnamefont {E.}~\bibnamefont
  {Lengyel}}, \bibinfo {author} {\bibfnamefont {M.}~\bibnamefont {Nicklas}},
  \bibinfo {author} {\bibfnamefont {N.}~\bibnamefont {Caroca-Canales}}, \ and\
  \bibinfo {author} {\bibfnamefont {C.}~\bibnamefont {Geibel}},\ }\bibfield
  {title} {\enquote {\bibinfo {title} {{Temperature-pressure phase diagram of
  CeCoSi: Pressure-induced high-temperature phase}},}\ }\href {\doibase
  10.1103/PhysRevB.88.155137} {\bibfield  {journal} {\bibinfo  {journal} {Phys.
  Rev. B}\ }\textbf {\bibinfo {volume} {88}},\ \bibinfo {pages} {155137}
  (\bibinfo {year} {2013})}\BibitemShut {NoStop}%
\bibitem [{\citenamefont {Tanida}\ \emph {et~al.}(2018)\citenamefont {Tanida},
  \citenamefont {Muro},\ and\ \citenamefont
  {Matsumura}}]{tanida2018substitution}%
  \BibitemOpen
  \bibfield  {author} {\bibinfo {author} {\bibfnamefont {H.}~\bibnamefont
  {Tanida}}, \bibinfo {author} {\bibfnamefont {Y.}~\bibnamefont {Muro}}, \ and\
  \bibinfo {author} {\bibfnamefont {T.}~\bibnamefont {Matsumura}},\ }\bibfield
  {title} {\enquote {\bibinfo {title} {{La Substitution and Pressure Studies on
  CeCoSi: A Possible Antiferroquadrupolar Ordering Induced by Pressure}},}\
  }\href {\doibase 10.7566/JPSJ.87.023705} {\bibfield  {journal} {\bibinfo
  {journal} {J. Phys. Soc. Jpn.}\ }\textbf {\bibinfo {volume} {87}},\ \bibinfo
  {pages} {023705} (\bibinfo {year} {2018})}\BibitemShut {NoStop}%
\bibitem [{\citenamefont {Chevalier}\ and\ \citenamefont
  {Matar}(2004)}]{chevalier2004effect}%
  \BibitemOpen
  \bibfield  {author} {\bibinfo {author} {\bibfnamefont {B.}~\bibnamefont
  {Chevalier}}\ and\ \bibinfo {author} {\bibfnamefont {S.~F.}\ \bibnamefont
  {Matar}},\ }\bibfield  {title} {\enquote {\bibinfo {title} {{Effect of H
  insertion on the magnetic, electronic, and structural properties of
  CeCoSi}},}\ }\href {\doibase 10.1103/PhysRevB.70.174408} {\bibfield
  {journal} {\bibinfo  {journal} {Phys. Rev. B}\ }\textbf {\bibinfo {volume}
  {70}},\ \bibinfo {pages} {174408} (\bibinfo {year} {2004})}\BibitemShut
  {NoStop}%
\bibitem [{\citenamefont {Chevalier}\ \emph {et~al.}(2006)\citenamefont
  {Chevalier}, \citenamefont {Matar}, \citenamefont {Marcos},\ and\
  \citenamefont {Fernandez}}]{chevalier2006antiferromagnetic}%
  \BibitemOpen
  \bibfield  {author} {\bibinfo {author} {\bibfnamefont {B.}~\bibnamefont
  {Chevalier}}, \bibinfo {author} {\bibfnamefont {S.~F.}\ \bibnamefont
  {Matar}}, \bibinfo {author} {\bibfnamefont {J.~S.}\ \bibnamefont {Marcos}}, \
  and\ \bibinfo {author} {\bibfnamefont {J.~R.}\ \bibnamefont {Fernandez}},\
  }\bibfield  {title} {\enquote {\bibinfo {title} {{From antiferromagnetic
  ordering to spin fluctuation behavior induced by hydrogenation of ternary
  compounds CeCoSi and CeCoGe}},}\ }\href {\doibase
  10.1016/j.physb.2006.01.291} {\bibfield  {journal} {\bibinfo  {journal}
  {Physica B}\ }\textbf {\bibinfo {volume} {378}},\ \bibinfo {pages} {795--796}
  (\bibinfo {year} {2006})}\BibitemShut {NoStop}%
\bibitem [{\citenamefont {Chevalier}\ and\ \citenamefont
  {Malaman}(2004)}]{chevalier2004reinvestigation}%
  \BibitemOpen
  \bibfield  {author} {\bibinfo {author} {\bibfnamefont {B.}~\bibnamefont
  {Chevalier}}\ and\ \bibinfo {author} {\bibfnamefont {B.}~\bibnamefont
  {Malaman}},\ }\bibfield  {title} {\enquote {\bibinfo {title}
  {{Reinvestigation of the electrical and magnetic properties of the ternary
  germanide CeCoGe}},}\ }\href {\doibase 10.1016/j.ssc.2004.04.010} {\bibfield
  {journal} {\bibinfo  {journal} {Solid State Commun.}\ }\textbf {\bibinfo
  {volume} {130}},\ \bibinfo {pages} {711--715} (\bibinfo {year}
  {2004})}\BibitemShut {NoStop}%
\bibitem [{\citenamefont {Tanida}\ \emph {et~al.}(2019)\citenamefont {Tanida},
  \citenamefont {Mitsumoto}, \citenamefont {Muro}, \citenamefont {Fukuhara},
  \citenamefont {Kawamura}, \citenamefont {Kondo}, \citenamefont {Kindo},
  \citenamefont {Matsumoto}, \citenamefont {Namiki}, \citenamefont {Kuwai},\
  and\ \citenamefont {Matsumura}}]{tanida2019successive}%
  \BibitemOpen
  \bibfield  {author} {\bibinfo {author} {\bibfnamefont {H.}~\bibnamefont
  {Tanida}}, \bibinfo {author} {\bibfnamefont {K.}~\bibnamefont {Mitsumoto}},
  \bibinfo {author} {\bibfnamefont {Y.}~\bibnamefont {Muro}}, \bibinfo {author}
  {\bibfnamefont {T.}~\bibnamefont {Fukuhara}}, \bibinfo {author}
  {\bibfnamefont {Y.}~\bibnamefont {Kawamura}}, \bibinfo {author}
  {\bibfnamefont {A.}~\bibnamefont {Kondo}}, \bibinfo {author} {\bibfnamefont
  {K.}~\bibnamefont {Kindo}}, \bibinfo {author} {\bibfnamefont
  {Y.}~\bibnamefont {Matsumoto}}, \bibinfo {author} {\bibfnamefont
  {T.}~\bibnamefont {Namiki}}, \bibinfo {author} {\bibfnamefont
  {T.}~\bibnamefont {Kuwai}}, \ and\ \bibinfo {author} {\bibfnamefont
  {T.}~\bibnamefont {Matsumura}},\ }\bibfield  {title} {\enquote {\bibinfo
  {title} {{Successive Phase Transition at Ambient Pressure in CeCoSi: Single
  Crystal Studies}},}\ }\href {\doibase 10.7566/JPSJ.88.054716} {\bibfield
  {journal} {\bibinfo  {journal} {J. Phys. Soc. Jpn.}\ }\textbf {\bibinfo
  {volume} {88}},\ \bibinfo {pages} {054716} (\bibinfo {year}
  {2019})}\BibitemShut {NoStop}%
\bibitem [{\citenamefont {M.~Manago}\ and\ \citenamefont
  {Tanida}(2019)}]{Manago2019}%
  \BibitemOpen
  \bibfield  {author} {\bibinfo {author} {\bibfnamefont {H.~Tou}\ \bibnamefont
  {M.~Manago}, \bibfnamefont {H.~Kotegawa}}\ and\ \bibinfo {author}
  {\bibfnamefont {H.}~\bibnamefont {Tanida}},\ }\bibfield  {title} {\enquote
  {\bibinfo {title} {{NMR evidence of a non-magnetic phase transition in
  CeCoSi}},}\ }\href
  {https://www.jphysics.jp/activity/2020/01/06/jphysics_FY2019_abstratcts_20200106.pdf}
  {\bibfield  {journal} {\bibinfo  {journal} {J-Physics Annual Meeting FY2019}\
  ,\ \bibinfo {pages} {P05}} (\bibinfo {year} {2019})}\BibitemShut {NoStop}%
\bibitem [{\citenamefont {Hattori}(2010)}]{hattori2010meta}%
  \BibitemOpen
  \bibfield  {author} {\bibinfo {author} {\bibfnamefont {K.}~\bibnamefont
  {Hattori}},\ }\bibfield  {title} {\enquote {\bibinfo {title} {{Meta-orbital
  transition in heavy-fermion systems: Analysis by dynamical mean field theory
  and self-consistent renormalization theory of orbital fluctuations}},}\
  }\href {\doibase 10.1143/JPSJ.79.114717} {\bibfield  {journal} {\bibinfo
  {journal} {J. Phys. Soc. Jpn.}\ }\textbf {\bibinfo {volume} {79}},\ \bibinfo
  {pages} {114717} (\bibinfo {year} {2010})}\BibitemShut {NoStop}%
\bibitem [{\citenamefont {Pourovskii}\ \emph {et~al.}(2014)\citenamefont
  {Pourovskii}, \citenamefont {Hansmann}, \citenamefont {Ferrero},\ and\
  \citenamefont {Georges}}]{pourovskii2014theoretical}%
  \BibitemOpen
  \bibfield  {author} {\bibinfo {author} {\bibfnamefont {L.~V.}\ \bibnamefont
  {Pourovskii}}, \bibinfo {author} {\bibfnamefont {P.}~\bibnamefont
  {Hansmann}}, \bibinfo {author} {\bibfnamefont {M.}~\bibnamefont {Ferrero}}, \
  and\ \bibinfo {author} {\bibfnamefont {A.}~\bibnamefont {Georges}},\
  }\bibfield  {title} {\enquote {\bibinfo {title} {{Theoretical prediction and
  spectroscopic fingerprints of an orbital transition in CeCu$_2$Si$_2$}},}\
  }\href {\doibase 10.1103/PhysRevLett.112.106407} {\bibfield  {journal}
  {\bibinfo  {journal} {Phys. Rev. Lett.}\ }\textbf {\bibinfo {volume} {112}},\
  \bibinfo {pages} {106407} (\bibinfo {year} {2014})}\BibitemShut {NoStop}%
\bibitem [{\citenamefont {Bewley}\ \emph {et~al.}(2011)\citenamefont {Bewley},
  \citenamefont {Taylor},\ and\ \citenamefont {Bennington}}]{bewley2011let}%
  \BibitemOpen
  \bibfield  {author} {\bibinfo {author} {\bibfnamefont {R.~I.}\ \bibnamefont
  {Bewley}}, \bibinfo {author} {\bibfnamefont {J.~W.}\ \bibnamefont {Taylor}},
  \ and\ \bibinfo {author} {\bibfnamefont {S.~M.}\ \bibnamefont {Bennington}},\
  }\bibfield  {title} {\enquote {\bibinfo {title} {{LET, a cold neutron
  multi-disk chopper spectrometer at ISIS}},}\ }\href {\doibase
  10.1016/j.nima.2011.01.173} {\bibfield  {journal} {\bibinfo  {journal} {Nucl.
  Instrum. Methods Phys. Res.}\ }\textbf {\bibinfo {volume} {637}},\ \bibinfo
  {pages} {128--134} (\bibinfo {year} {2011})}\BibitemShut {NoStop}%
\bibitem [{\citenamefont {Ehlers}\ \emph {et~al.}(2011)\citenamefont {Ehlers},
  \citenamefont {Podlesnyak}, \citenamefont {Niedziela}, \citenamefont
  {Iverson},\ and\ \citenamefont {Sokol}}]{CNCS1}%
  \BibitemOpen
  \bibfield  {author} {\bibinfo {author} {\bibfnamefont {G.}~\bibnamefont
  {Ehlers}}, \bibinfo {author} {\bibfnamefont {A.}~\bibnamefont {Podlesnyak}},
  \bibinfo {author} {\bibfnamefont {J.~L.}\ \bibnamefont {Niedziela}}, \bibinfo
  {author} {\bibfnamefont {E.~B.}\ \bibnamefont {Iverson}}, \ and\ \bibinfo
  {author} {\bibfnamefont {P.~E.}\ \bibnamefont {Sokol}},\ }\bibfield  {title}
  {\enquote {\bibinfo {title} {{The new cold neutron chopper spectrometer at
  the Spallation Neutron Source: design and performance}},}\ }\href {\doibase
  10.1063/1.3626935} {\bibfield  {journal} {\bibinfo  {journal} {Rev. Sci.
  Instrum.}\ }\textbf {\bibinfo {volume} {82}},\ \bibinfo {pages} {085108}
  (\bibinfo {year} {2011})}\BibitemShut {NoStop}%
\bibitem [{\citenamefont {Ehlers}\ \emph {et~al.}(2016)\citenamefont {Ehlers},
  \citenamefont {Podlesnyak},\ and\ \citenamefont {Kolesnikov}}]{CNCS2}%
  \BibitemOpen
  \bibfield  {author} {\bibinfo {author} {\bibfnamefont {G.}~\bibnamefont
  {Ehlers}}, \bibinfo {author} {\bibfnamefont {A.}~\bibnamefont {Podlesnyak}},
  \ and\ \bibinfo {author} {\bibfnamefont {A.~I.}\ \bibnamefont {Kolesnikov}},\
  }\bibfield  {title} {\enquote {\bibinfo {title} {{The cold neutron chopper
  spectrometer at the Spallation Neutron Source - A review of the first 8 years
  of operation}},}\ }\href {\doibase 10.1063/1.4962024} {\bibfield  {journal}
  {\bibinfo  {journal} {Rev. Sci. Instrum.}\ }\textbf {\bibinfo {volume}
  {87}},\ \bibinfo {pages} {093902} (\bibinfo {year} {2016})}\BibitemShut
  {NoStop}%
\bibitem [{\citenamefont {{R. A. Forman and G. J. Piermarini and J. D. Barnett
  and S. Block}}(1972)}]{forman1972pressure}%
  \BibitemOpen
  \bibfield  {author} {\bibinfo {author} {\bibnamefont {{R. A. Forman and G. J.
  Piermarini and J. D. Barnett and S. Block}}},\ }\bibfield  {title} {\enquote
  {\bibinfo {title} {Pressure measurement made by the utilization of ruby
  sharp-line luminescence},}\ }\href {\doibase 10.1126/science.176.4032.284}
  {\bibfield  {journal} {\bibinfo  {journal} {Science}\ }\textbf {\bibinfo
  {volume} {176}},\ \bibinfo {pages} {284--285} (\bibinfo {year}
  {1972})}\BibitemShut {NoStop}%
\bibitem [{\citenamefont {Podlesnyak}\ \emph {et~al.}(2018)\citenamefont
  {Podlesnyak}, \citenamefont {Loguillo}, \citenamefont {Rucker}, \citenamefont
  {Haberl}, \citenamefont {Boehler}, \citenamefont {Ehlers}, \citenamefont
  {Daemen}, \citenamefont {Armitage}, \citenamefont {Frontzek},\ and\
  \citenamefont {Lumsden}}]{podlesnyak2018clamp}%
  \BibitemOpen
  \bibfield  {author} {\bibinfo {author} {\bibfnamefont {A.}~\bibnamefont
  {Podlesnyak}}, \bibinfo {author} {\bibfnamefont {M.}~\bibnamefont
  {Loguillo}}, \bibinfo {author} {\bibfnamefont {G.~M.}\ \bibnamefont
  {Rucker}}, \bibinfo {author} {\bibfnamefont {B.}~\bibnamefont {Haberl}},
  \bibinfo {author} {\bibfnamefont {R.}~\bibnamefont {Boehler}}, \bibinfo
  {author} {\bibfnamefont {G.}~\bibnamefont {Ehlers}}, \bibinfo {author}
  {\bibfnamefont {L.~L.}\ \bibnamefont {Daemen}}, \bibinfo {author}
  {\bibfnamefont {D.}~\bibnamefont {Armitage}}, \bibinfo {author}
  {\bibfnamefont {M.~D.}\ \bibnamefont {Frontzek}}, \ and\ \bibinfo {author}
  {\bibfnamefont {M.}~\bibnamefont {Lumsden}},\ }\bibfield  {title} {\enquote
  {\bibinfo {title} {{Clamp cell with in situ pressure monitoring for
  low-temperature neutron scattering measurements}},}\ }\href {\doibase
  10.1080/08957959.2018.1519560} {\bibfield  {journal} {\bibinfo  {journal}
  {High Pressure Res.}\ }\textbf {\bibinfo {volume} {38}},\ \bibinfo {pages}
  {482--492} (\bibinfo {year} {2018})}\BibitemShut {NoStop}%
\bibitem [{Note1()}]{Note1}%
  \BibitemOpen
  \bibinfo {note} {Using of $E_{\protect \mathrm {i}} > 19$~meV could provide
  better energy range for studying the CEF excitations, but would strongly
  decrease the neutron flux on the sample.}\BibitemShut {Stop}%
\bibitem [{\citenamefont {Pet{\v{r}}{\'\i}{\v{c}}ek}\ \emph
  {et~al.}(2014)\citenamefont {Pet{\v{r}}{\'\i}{\v{c}}ek}, \citenamefont
  {Du{\v{s}}ek},\ and\ \citenamefont {Palatinus}}]{JANA}%
  \BibitemOpen
  \bibfield  {author} {\bibinfo {author} {\bibfnamefont {V.}~\bibnamefont
  {Pet{\v{r}}{\'\i}{\v{c}}ek}}, \bibinfo {author} {\bibfnamefont
  {M.}~\bibnamefont {Du{\v{s}}ek}}, \ and\ \bibinfo {author} {\bibfnamefont
  {L.}~\bibnamefont {Palatinus}},\ }\bibfield  {title} {\enquote {\bibinfo
  {title} {{Crystallographic computing system JANA2006: general features}},}\
  }\href {\doibase 10.1515/zkri-2014-1737} {\bibfield  {journal} {\bibinfo
  {journal} {Z. Kristallogr. Cryst. Mater.}\ }\textbf {\bibinfo {volume}
  {229}},\ \bibinfo {pages} {345--352} (\bibinfo {year} {2014})}\BibitemShut
  {NoStop}%
\bibitem [{\citenamefont {Azuah}\ \emph {et~al.}(2009)\citenamefont {Azuah},
  \citenamefont {Kneller}, \citenamefont {Qiu}, \citenamefont
  {Tregenna-Piggott}, \citenamefont {Brown}, \citenamefont {Copley},\ and\
  \citenamefont {Dimeo}}]{AzuahKneller09}%
  \BibitemOpen
  \bibfield  {author} {\bibinfo {author} {\bibfnamefont {R.~T.}\ \bibnamefont
  {Azuah}}, \bibinfo {author} {\bibfnamefont {L.~R.}\ \bibnamefont {Kneller}},
  \bibinfo {author} {\bibfnamefont {Y.}~\bibnamefont {Qiu}}, \bibinfo {author}
  {\bibfnamefont {P.~L.~W.}\ \bibnamefont {Tregenna-Piggott}}, \bibinfo
  {author} {\bibfnamefont {C.~M.}\ \bibnamefont {Brown}}, \bibinfo {author}
  {\bibfnamefont {J.~R.~D.}\ \bibnamefont {Copley}}, \ and\ \bibinfo {author}
  {\bibfnamefont {R.~M.}\ \bibnamefont {Dimeo}},\ }\bibfield  {title} {\enquote
  {\bibinfo {title} {{DAVE: a comprehensive software suite for the reduction,
  visualization, and analysis of low energy neutron spectroscopic data}},}\
  }\href {\doibase 10.6028/jres.114.025} {\bibfield  {journal} {\bibinfo
  {journal} {J. Res. Natl. Inst. Stan. Technol.}\ }\textbf {\bibinfo {volume}
  {114}},\ \bibinfo {pages} {341} (\bibinfo {year} {2009})}\BibitemShut
  {NoStop}%
\bibitem [{\citenamefont {Arnold}\ \emph {et~al.}(2014)\citenamefont {Arnold},
  \citenamefont {Bilheux}, \citenamefont {Borreguero}, \citenamefont {Buts},
  \citenamefont {Campbell}, \citenamefont {Chapon}, \citenamefont {Doucet},
  \citenamefont {Draper}, \citenamefont {Leal}, \citenamefont {Gigg},
  \citenamefont {Lynch}, \citenamefont {Markvardsen}, \citenamefont
  {Mikkelson}, \citenamefont {Mikkelson}, \citenamefont {Miller}, \citenamefont
  {Palmen}, \citenamefont {Parker}, \citenamefont {Passos}, \citenamefont
  {Perring}, \citenamefont {Peterson}, \citenamefont {Ren}, \citenamefont
  {Reuter}, \citenamefont {Savici}, \citenamefont {Taylor}, \citenamefont
  {Taylor}, \citenamefont {Tolchenov}, \citenamefont {Zhou},\ and\
  \citenamefont {Zikovsky}}]{Mantid}%
  \BibitemOpen
  \bibfield  {author} {\bibinfo {author} {\bibfnamefont {O.}~\bibnamefont
  {Arnold}}, \bibinfo {author} {\bibfnamefont {J.~C.}\ \bibnamefont {Bilheux}},
  \bibinfo {author} {\bibfnamefont {J.~M.}\ \bibnamefont {Borreguero}},
  \bibinfo {author} {\bibfnamefont {A.}~\bibnamefont {Buts}}, \bibinfo {author}
  {\bibfnamefont {S.~I.}\ \bibnamefont {Campbell}}, \bibinfo {author}
  {\bibfnamefont {L.}~\bibnamefont {Chapon}}, \bibinfo {author} {\bibfnamefont
  {M.}~\bibnamefont {Doucet}}, \bibinfo {author} {\bibfnamefont
  {N.}~\bibnamefont {Draper}}, \bibinfo {author} {\bibfnamefont {R.~Ferraz}\
  \bibnamefont {Leal}}, \bibinfo {author} {\bibfnamefont {M.~A.}\ \bibnamefont
  {Gigg}}, \bibinfo {author} {\bibfnamefont {V.~E.}\ \bibnamefont {Lynch}},
  \bibinfo {author} {\bibfnamefont {A.}~\bibnamefont {Markvardsen}}, \bibinfo
  {author} {\bibfnamefont {D.~J.}\ \bibnamefont {Mikkelson}}, \bibinfo {author}
  {\bibfnamefont {R.~L.}\ \bibnamefont {Mikkelson}}, \bibinfo {author}
  {\bibfnamefont {R.}~\bibnamefont {Miller}}, \bibinfo {author} {\bibfnamefont
  {K.}~\bibnamefont {Palmen}}, \bibinfo {author} {\bibfnamefont
  {P.}~\bibnamefont {Parker}}, \bibinfo {author} {\bibfnamefont
  {G.}~\bibnamefont {Passos}}, \bibinfo {author} {\bibfnamefont {T.~G.}\
  \bibnamefont {Perring}}, \bibinfo {author} {\bibfnamefont {P.~F.}\
  \bibnamefont {Peterson}}, \bibinfo {author} {\bibfnamefont {S.}~\bibnamefont
  {Ren}}, \bibinfo {author} {\bibfnamefont {M.~A.}\ \bibnamefont {Reuter}},
  \bibinfo {author} {\bibfnamefont {A.~T.}\ \bibnamefont {Savici}}, \bibinfo
  {author} {\bibfnamefont {J.~W.}\ \bibnamefont {Taylor}}, \bibinfo {author}
  {\bibfnamefont {R.~J.}\ \bibnamefont {Taylor}}, \bibinfo {author}
  {\bibfnamefont {R.}~\bibnamefont {Tolchenov}}, \bibinfo {author}
  {\bibfnamefont {W.}~\bibnamefont {Zhou}}, \ and\ \bibinfo {author}
  {\bibfnamefont {J.}~\bibnamefont {Zikovsky}},\ }\bibfield  {title} {\enquote
  {\bibinfo {title} {{Mantid -- Data analysis and visualization package for
  neutron scattering and $\mu$SR experiments}},}\ }\href {\doibase
  10.1016/j.nima.2014.07.029} {\bibfield  {journal} {\bibinfo  {journal} {Nucl.
  Instrum. Methods Phys. Res. Sect. A}\ }\textbf {\bibinfo {volume} {764}},\
  \bibinfo {pages} {156} (\bibinfo {year} {2014})}\BibitemShut {NoStop}%
\bibitem [{\citenamefont {Richard}\ \emph {et~al.}(1996)\citenamefont
  {Richard}, \citenamefont {Ferrand},\ and\ \citenamefont {Kearley}}]{lamp}%
  \BibitemOpen
  \bibfield  {author} {\bibinfo {author} {\bibfnamefont {D.}~\bibnamefont
  {Richard}}, \bibinfo {author} {\bibfnamefont {M.}~\bibnamefont {Ferrand}}, \
  and\ \bibinfo {author} {\bibfnamefont {G.~J.}\ \bibnamefont {Kearley}},\
  }\bibfield  {title} {\enquote {\bibinfo {title} {{Analysis and visualisation
  of neutron-scattering data}},}\ }\href {\doibase 10.1080/10238169608200065}
  {\bibfield  {journal} {\bibinfo  {journal} {J. Neutron Res.}\ }\textbf
  {\bibinfo {volume} {4}},\ \bibinfo {pages} {33--39} (\bibinfo {year}
  {1996})}\BibitemShut {NoStop}%
\bibitem [{Note2()}]{Note2}%
  \BibitemOpen
  \bibinfo {note} {We also measured spectra with higher $E_{\protect \mathrm
  {i}} = 67.6$~meV, but no additional magnetic excitations were observed in the
  spectra.}\BibitemShut {Stop}%
\bibitem [{Note3()}]{Note3}%
  \BibitemOpen
  \bibinfo {note} {We also tried a classical approach proposed by A. P. Murani
  for CeSn$_3$~\cite {murani1983magnetic}, and the results are essentially
  identical.}\BibitemShut {Stop}%
\bibitem [{\citenamefont {{\v{C}}erm{\'a}k}\ \emph {et~al.}(2019)\citenamefont
  {{\v{C}}erm{\'a}k}, \citenamefont {Schneidewind}, \citenamefont {Liu},
  \citenamefont {Koza}, \citenamefont {Franz}, \citenamefont {Sch{\"o}nmann},
  \citenamefont {Sobolev},\ and\ \citenamefont
  {Pfleiderer}}]{vcermak2019magnetoelastic}%
  \BibitemOpen
  \bibfield  {author} {\bibinfo {author} {\bibfnamefont {Petr}\ \bibnamefont
  {{\v{C}}erm{\'a}k}}, \bibinfo {author} {\bibfnamefont {Astrid}\ \bibnamefont
  {Schneidewind}}, \bibinfo {author} {\bibfnamefont {Benqiong}\ \bibnamefont
  {Liu}}, \bibinfo {author} {\bibfnamefont {Michael~Marek}\ \bibnamefont
  {Koza}}, \bibinfo {author} {\bibfnamefont {Christian}\ \bibnamefont {Franz}},
  \bibinfo {author} {\bibfnamefont {Rudolf}\ \bibnamefont {Sch{\"o}nmann}},
  \bibinfo {author} {\bibfnamefont {Oleg}\ \bibnamefont {Sobolev}}, \ and\
  \bibinfo {author} {\bibfnamefont {Christian}\ \bibnamefont {Pfleiderer}},\
  }\bibfield  {title} {\enquote {\bibinfo {title} {{Magnetoelastic hybrid
  excitations in CeAuAl$_3$}},}\ }\href {\doibase 10.1073/pnas.1819664116}
  {\bibfield  {journal} {\bibinfo  {journal} {Proc. Natl. Acad. Sci.}\ }\textbf
  {\bibinfo {volume} {116}},\ \bibinfo {pages} {6695--6700} (\bibinfo {year}
  {2019})}\BibitemShut {NoStop}%
\bibitem [{\citenamefont {Continentino}\ \emph {et~al.}(2001)\citenamefont
  {Continentino}, \citenamefont {de~Medeiros}, \citenamefont {Orlando},
  \citenamefont {Fontes},\ and\ \citenamefont
  {Baggio-Saitovitch}}]{continentino2001anisotropic}%
  \BibitemOpen
  \bibfield  {author} {\bibinfo {author} {\bibfnamefont {M.~A.}\ \bibnamefont
  {Continentino}}, \bibinfo {author} {\bibfnamefont {S.~N.}\ \bibnamefont
  {de~Medeiros}}, \bibinfo {author} {\bibfnamefont {M.~T.~D.}\ \bibnamefont
  {Orlando}}, \bibinfo {author} {\bibfnamefont {M.~B.}\ \bibnamefont {Fontes}},
  \ and\ \bibinfo {author} {\bibfnamefont {E.~M.}\ \bibnamefont
  {Baggio-Saitovitch}},\ }\bibfield  {title} {\enquote {\bibinfo {title}
  {{Anisotropic quantum critical behavior in CeCoGe$_{3-x}$Si$_x$}},}\ }\href
  {\doibase 10.1103/PhysRevB.64.012404} {\bibfield  {journal} {\bibinfo
  {journal} {Phys. Rev. B}\ }\textbf {\bibinfo {volume} {64}},\ \bibinfo
  {pages} {012404} (\bibinfo {year} {2001})}\BibitemShut {NoStop}%
\bibitem [{Note4()}]{Note4}%
  \BibitemOpen
  \bibinfo {note} {Note that the signal at 0.6~GPa was measured during a
  separate beamtime with slightly different setup, which did not allowed us to
  subtract the background precisely. This problem causes sharp features in
  Fig.~\ref {CNCS}(a) and is responsible for the lower intensity of 0.6~GPa
  signal shown in the Fig.~\ref {CNCS}(b).}\BibitemShut {Stop}%
\bibitem [{Note5()}]{Note5}%
  \BibitemOpen
  \bibinfo {note} {Investigation of the form-factor of the excitation is not
  possible due to the cell phonon scattering at high $\protect \mathbf {Q}$
  above $\sim $3~\r A$^{-1}$.}\BibitemShut {Stop}%
\bibitem [{\citenamefont {Fischer}\ and\ \citenamefont
  {Herr}(1990)}]{fischer1990mean}%
  \BibitemOpen
  \bibfield  {author} {\bibinfo {author} {\bibfnamefont {G.}~\bibnamefont
  {Fischer}}\ and\ \bibinfo {author} {\bibfnamefont {A.}~\bibnamefont {Herr}},\
  }\bibfield  {title} {\enquote {\bibinfo {title} {{Mean magnetic moments of
  polycrystalline Ce compounds in a tetragonal crystal field}},}\ }\href
  {https://onlinelibrary.wiley.com/doi/pdf/10.1002/pssb.2221590161} {\bibfield
  {journal} {\bibinfo  {journal} {Phys. Stat. Sol. (B)}\ }\textbf {\bibinfo
  {volume} {159}},\ \bibinfo {pages} {K23--K26} (\bibinfo {year}
  {1990})}\BibitemShut {NoStop}%
\bibitem [{\citenamefont {Besnus}\ \emph {et~al.}(1992)\citenamefont {Besnus},
  \citenamefont {Braghta}, \citenamefont {Hamdaoui},\ and\ \citenamefont
  {Meyer}}]{besnus1992correlation}%
  \BibitemOpen
  \bibfield  {author} {\bibinfo {author} {\bibfnamefont {M.~J.}\ \bibnamefont
  {Besnus}}, \bibinfo {author} {\bibfnamefont {A.}~\bibnamefont {Braghta}},
  \bibinfo {author} {\bibfnamefont {N.}~\bibnamefont {Hamdaoui}}, \ and\
  \bibinfo {author} {\bibfnamefont {A.}~\bibnamefont {Meyer}},\ }\bibfield
  {title} {\enquote {\bibinfo {title} {{A correlation between specific heat and
  the ratio $T_{\mathrm{K}}/T_{\mathrm{N}}$ in magnetic Kondo lattices}},}\
  }\href {\doibase 10.1016/0304-8853(92)90630-7} {\bibfield  {journal}
  {\bibinfo  {journal} {J. Mag. Mag. Mat.}\ }\textbf {\bibinfo {volume}
  {104}},\ \bibinfo {pages} {1385--1386} (\bibinfo {year} {1992})}\BibitemShut
  {NoStop}%
\bibitem [{\citenamefont {Bredl}\ \emph {et~al.}(1978)\citenamefont {Bredl},
  \citenamefont {Steglich},\ and\ \citenamefont {Schotte}}]{bredl1978specific}%
  \BibitemOpen
  \bibfield  {author} {\bibinfo {author} {\bibfnamefont {C.~D.}\ \bibnamefont
  {Bredl}}, \bibinfo {author} {\bibfnamefont {F.}~\bibnamefont {Steglich}}, \
  and\ \bibinfo {author} {\bibfnamefont {K.~D.}\ \bibnamefont {Schotte}},\
  }\bibfield  {title} {\enquote {\bibinfo {title} {{Specific heat of
  concentrated kondo systems: (La,Ce)Al$_2$ and CeAl$_2$}},}\ }\href {\doibase
  10.1007/BF01324030} {\bibfield  {journal} {\bibinfo  {journal} {Z. Physik B}\
  }\textbf {\bibinfo {volume} {29}},\ \bibinfo {pages} {327--340} (\bibinfo
  {year} {1978})}\BibitemShut {NoStop}%
\bibitem [{\citenamefont {Kr{\"u}ger}\ \emph {et~al.}(2014)\citenamefont
  {Kr{\"u}ger}, \citenamefont {Pedder},\ and\ \citenamefont
  {Green}}]{kruger2014fluctuation}%
  \BibitemOpen
  \bibfield  {author} {\bibinfo {author} {\bibfnamefont {F.}~\bibnamefont
  {Kr{\"u}ger}}, \bibinfo {author} {\bibfnamefont {C.~J.}\ \bibnamefont
  {Pedder}}, \ and\ \bibinfo {author} {\bibfnamefont {A.~G.}\ \bibnamefont
  {Green}},\ }\bibfield  {title} {\enquote {\bibinfo {title}
  {{Fluctuation-driven magnetic hard-axis ordering in metallic
  ferromagnets}},}\ }\href {\doibase 10.1103/PhysRevLett.113.147001} {\bibfield
   {journal} {\bibinfo  {journal} {Phys. Rev. Lett.}\ }\textbf {\bibinfo
  {volume} {113}},\ \bibinfo {pages} {147001} (\bibinfo {year}
  {2014})}\BibitemShut {NoStop}%
\bibitem [{\citenamefont {Hafner}\ \emph {et~al.}(2019)\citenamefont {Hafner},
  \citenamefont {Rai}, \citenamefont {Banda}, \citenamefont {Kliemt},
  \citenamefont {Krellner}, \citenamefont {Sichelschmidt}, \citenamefont
  {Morosan}, \citenamefont {Geibel},\ and\ \citenamefont
  {Brando}}]{hafner2019kondo}%
  \BibitemOpen
  \bibfield  {author} {\bibinfo {author} {\bibfnamefont {D.}~\bibnamefont
  {Hafner}}, \bibinfo {author} {\bibfnamefont {B.~K.}\ \bibnamefont {Rai}},
  \bibinfo {author} {\bibfnamefont {J.}~\bibnamefont {Banda}}, \bibinfo
  {author} {\bibfnamefont {K.}~\bibnamefont {Kliemt}}, \bibinfo {author}
  {\bibfnamefont {C.}~\bibnamefont {Krellner}}, \bibinfo {author}
  {\bibfnamefont {J.}~\bibnamefont {Sichelschmidt}}, \bibinfo {author}
  {\bibfnamefont {E.}~\bibnamefont {Morosan}}, \bibinfo {author} {\bibfnamefont
  {C.}~\bibnamefont {Geibel}}, \ and\ \bibinfo {author} {\bibfnamefont
  {M.}~\bibnamefont {Brando}},\ }\bibfield  {title} {\enquote {\bibinfo {title}
  {{Kondo-lattice ferromagnets and their peculiar order along the magnetically
  hard axis determined by the crystalline electric field}},}\ }\href {\doibase
  10.1103/PhysRevB.99.201109} {\bibfield  {journal} {\bibinfo  {journal} {Phys.
  Rev. B}\ }\textbf {\bibinfo {volume} {99}},\ \bibinfo {pages} {201109}
  (\bibinfo {year} {2019})}\BibitemShut {NoStop}%
\bibitem [{\citenamefont {Cooper}(1972)}]{Cooper1972Magnetic}%
  \BibitemOpen
  \bibfield  {author} {\bibinfo {author} {\bibfnamefont {B.~R.}\ \bibnamefont
  {Cooper}},\ }\enquote {\bibinfo {title} {Magnetic properties of rare earth
  metals},}\ \ (\bibinfo  {publisher} {Springer, Boston},\ \bibinfo {year}
  {1972})\ pp.\ \bibinfo {pages} {17--80}\BibitemShut {NoStop}%
\bibitem [{\citenamefont {Pedregosa}\ \emph {et~al.}(2011)\citenamefont
  {Pedregosa}, \citenamefont {Varoquaux}, \citenamefont {Gramfort},
  \citenamefont {Michel}, \citenamefont {Thirion}, \citenamefont {Grisel},
  \citenamefont {Blondel}, \citenamefont {Prettenhofer}, \citenamefont {Weiss},
  \citenamefont {Dubourg}, \citenamefont {Vanderplas}, \citenamefont {Passos},
  \citenamefont {Cournapeau}, \citenamefont {Brucher}, \citenamefont {Perrot},\
  and\ \citenamefont {Duchesnay}}]{scikit}%
  \BibitemOpen
  \bibfield  {author} {\bibinfo {author} {\bibfnamefont {F.}~\bibnamefont
  {Pedregosa}}, \bibinfo {author} {\bibfnamefont {G.}~\bibnamefont
  {Varoquaux}}, \bibinfo {author} {\bibfnamefont {A.}~\bibnamefont {Gramfort}},
  \bibinfo {author} {\bibfnamefont {V.}~\bibnamefont {Michel}}, \bibinfo
  {author} {\bibfnamefont {B.}~\bibnamefont {Thirion}}, \bibinfo {author}
  {\bibfnamefont {O.}~\bibnamefont {Grisel}}, \bibinfo {author} {\bibfnamefont
  {M.}~\bibnamefont {Blondel}}, \bibinfo {author} {\bibfnamefont
  {P.}~\bibnamefont {Prettenhofer}}, \bibinfo {author} {\bibfnamefont
  {R.}~\bibnamefont {Weiss}}, \bibinfo {author} {\bibfnamefont
  {V.}~\bibnamefont {Dubourg}}, \bibinfo {author} {\bibfnamefont
  {J.}~\bibnamefont {Vanderplas}}, \bibinfo {author} {\bibfnamefont
  {A.}~\bibnamefont {Passos}}, \bibinfo {author} {\bibfnamefont
  {D.}~\bibnamefont {Cournapeau}}, \bibinfo {author} {\bibfnamefont
  {M.}~\bibnamefont {Brucher}}, \bibinfo {author} {\bibfnamefont
  {M.}~\bibnamefont {Perrot}}, \ and\ \bibinfo {author} {\bibfnamefont
  {E.}~\bibnamefont {Duchesnay}},\ }\bibfield  {title} {\enquote {\bibinfo
  {title} {Scikit-learn: Machine learning in {P}ython},}\ }\href
  {http://www.jmlr.org/papers/volume12/pedregosa11a/pedregosa11a.pdf}
  {\bibfield  {journal} {\bibinfo  {journal} {J. Mach. Learn. Res.}\ }\textbf
  {\bibinfo {volume} {12}},\ \bibinfo {pages} {2825--2830} (\bibinfo {year}
  {2011})}\BibitemShut {NoStop}%
\bibitem [{\citenamefont {Murani}(1983)}]{murani1983magnetic}%
  \BibitemOpen
  \bibfield  {author} {\bibinfo {author} {\bibfnamefont {A.~P.}\ \bibnamefont
  {Murani}},\ }\bibfield  {title} {\enquote {\bibinfo {title} {{Magnetic
  spectral response in the intermetallic compound CeSn$_3$}},}\ }\href
  {\doibase 10.1103/PhysRevB.28.2308} {\bibfield  {journal} {\bibinfo
  {journal} {Phys. Rev. B}\ }\textbf {\bibinfo {volume} {28}},\ \bibinfo
  {pages} {2308} (\bibinfo {year} {1983})}\BibitemShut {NoStop}%
\end{thebibliography}%

\end{document}